\documentstyle[
preprint,
prd,aps]{revtex}
\begin{document}
\noindent 
\hspace*{64ex} OCU-PHUS 179 (Revised) \\
\hspace*{64ex} October 2001
\begin{center} 
{\LARGE Fermion propagator in out of equilibrium quantum-field} \\
{\LARGE system and the Boltzmann equation} 
\end{center} 
\begin{center} 
{\large A. Ni\'{e}gawa}\footnote{Electronic address: 
niegawa@sci.osaka-cu.ac.jp} \\ 
{\it Department of Physics, Osaka City University, 
Sumiyoshi-ku, Osaka 558-8585, JAPAN}
\end{center} 
\hspace*{4ex} 
\begin{center} 
{\large Abstract} 
\end{center} 
We aim to construct from first principles a perturbative framework 
for studying nonequilibrium quantum-field systems that include 
massless Dirac fermions. The system of our concern is quasiuniform 
system near equilibrium or nonequilibrium quasistationary system. We 
employ the closed-time-path formalism and use the so-called gradient 
approximation. Essentially no further approximation is introduced. 
We construct a fermion propagator, with which a well-defined 
perturbative framework is formulated. In the course of construction 
of the framework, we obtain the generalized Boltzmann equation (GBE) 
that describes the evolution of the number-density functions of 
(anti)fermionic quasiparticles. 

\hspace*{1ex} 

11.10.Wx, 12.38.Mh, 12.38.Bx 
%%%%%%%%%%%%%%%%%%%%%%%%%%%%%%%%%%%%%%%%%%%%%%
%%%% SECTION I %%%%%%%%%%%%%%%%%%%%%%%%%%%%%%%
%%%%%%%%%%%%%%%%%%%%%%%%%%%%%%%%%%%%%%%%%%%%%%

\hspace*{1ex} 

\hspace*{1ex}  
\narrowtext 
\setcounter{equation}{0}
\setcounter{section}{0}
\def\theequation{\mbox{\arabic{section}.\arabic{equation}}} 
\section{Introduction} 
Ultrarelativistic heavy-ion-collision experiments at the BNL 
Relativistic Heavy Ion Collider (RHIC) have begun and will soon 
start at the CERN Large Hadron Collider (LHC) in an anticipation of 
producing a quark-gluon plasma (QGP) [see, e.g., \cite{bjo,le-b}]. 
The QGP to be produced is an 
expanding nonequilibrium system. Studies of the QGP as such have 
just begun. 

In previous papers, a perturbative framework has been formulated 
{\em from first principles} for dealing with out-of-equilibrium 
complex-scalar field system \cite{nie}, $O(N)$ linear-sigma system 
\cite{nie1}, and the system that includes gauge bosons \cite{nie2}. 
In this paper, we take up the out-of-equilibrium quantum-field 
theories that include Dirac fermions. Keeping in mind that the 
masses of light quarks may be ignored in QGP, we treat massless 
fermions. [Generalization to massive ones is straightforward.] 
Through similar procedure as in \cite{nie,nie1,nie2}, we construct 
the fermion propagator and, thereby, frame a perturbation theory. 
Essentially, only approximation we employ is the so-called gradient 
approximation (see below). We use the closed-time-path (CTP) 
formalism \cite{sch,chou,lan} of nonequilibrium statistical 
quantum-field theory. 

Throughout this paper, we are interested in quasiuniform systems 
near equilibrium or nonequilibrium quasistationary systems. Such 
systems are characterized (cf., e.g., \cite{ch}) by two different 
spacetime scales: microscopic or quantum-field-theoretical and 
macroscopic or statistical. The first scale, the 
microscopic-correlation scale, characterizes the range of radiative 
correction to reactions taking place in the system while the second 
scale measures the relaxation of the system. For a weak-coupling 
theory, in which we are interested in this paper, the former scale 
is much smaller than the latter scale. A well-known intuitive 
picture \cite{ch} for dealing with such systems is to separate 
spacetime into many \lq\lq cells'' whose characteristic size, 
$L^\mu$ $(\mu = 0, ..., 3)$, is in between the microscopic and 
macroscopic scales. It is assumed that the correlation between 
different cells is small, so that microscopic or elementary 
reactions can be regarded, to a good approximation, as taking place 
in a single cell. On the other hand, in a single cell, prominent 
relaxation phenomena do not take place. This intuitive picture may 
be implemented as follows. Let $S (x, y)$ be a propagator. For a 
system of our concern, $S (x, y)$, with $x^\mu - y^\mu$ [$|x^\mu - 
y^\mu| \lesssim L^\mu$] fixed, does not change appreciably within a 
single cell. Thus, the mid-point $X^\mu \equiv (x^\mu + y^\mu) / 2$ 
may be used as a label for the spacetime cells and is called the 
macroscopic spacetime coordinates or a slow variable \cite{lm}. On 
the other hand, relative spacetime coordinates $x^\mu - y^\mu$, 
which is called a fast variable, are responsible for describing 
microscopic reactions taking place in a single spacetime cell. We 
introduce a Wigner transformation (Fourier transformation with 
respect to the relative coordinates $x - y$ with $(x + y) / 2$ 
held fixed): 
%%%%%%%%%%%%%%%%%%%%%%%%%%%%%%%%%%%%%%%%%%%%
\begin{equation} 
S (x, y) = \int \frac{d^{\, 4} P}{(2 \pi)^4} \, e^{- i P \cdot (x 
- y)} \, S (X; P) 
\label{Fou} 
\end{equation} 
%%%%%%%%%%%%%%%%%%%%%%%%%
(with $P^\mu = (p^0, {\bf p})$). The Wigner function $S (X; P)$ 
depends on $X$ only weakly. The self-energy part $\Sigma (x, y)$ 
enjoys a similar property. We shall freely use $S(x, y)$ or $S(X; 
P)$ [$\Sigma (x, y)$ or $\Sigma (X; P)$], which we simply write $S$ 
[$\Sigma$] whenever obvious from the context. 

In the following, there often appears a \lq\lq product'' of 
functions: 
%%%%%%%%%%%%%%%%%%%%%%%
\begin{equation} 
[F \cdot G] (x, y) = \int d^{\, 4} z \, F (x, z) G (z, y) , 
\label{seki} 
\end{equation} 
%%%%%%%%%%%%%%%%%%%%%%%%%%%%%%
For the Wigner transform of the function $[F \cdot G] (x, y)$, we 
use the gradient 
approximation, 
%%%%%%%%%%%%%%%%%%%%%%%%%%%%%%%%%%%%%%%%%%%%%%%%%%%%%%%%%%
\begin{eqnarray} 
[F \cdot G] (X; P) & = & \int d^{\, 4} (x - y) \, e^{i P \cdot 
(x - y)} [F \cdot G] (x, y) \nonumber \\ 
& \simeq & F (X; P) G (X; P) \nonumber \\ 
&& - \frac{i}{2} \left\{ F (X; P) , \; G 
(X; P) \right\}_{\mbox{\scriptsize{P.B.}}} , 
\label{gra} 
\end{eqnarray} 
%%%%%%%%%%%%%%%%%%%%%%%%%%%%%%%%%%%%%%%%%%%%%
where $X = (x + y) / 2$ and 
%%%%%%%%%%%%%%%%%%%%%%%%%%%%%%%%%%%%%%
\begin{eqnarray} 
\left\{ F , \; G \right\}_{\mbox{\scriptsize{P.B.}}} & \equiv & 
\frac{\partial F (X; P)}{\partial X^\mu} \frac{\partial G (X; 
P)}{\partial P_\mu} \nonumber \\ 
&& - \frac{\partial F (X; P)}{\partial P_\mu} 
\frac{\partial G (X; P)}{\partial X^\mu} . 
\label{poi} 
\end{eqnarray} 
%%%%%%%%%%%%%%%%%%%%%%%%%%%%%%%%%%%%%%%%%%%%%%%%
We refer to the first term on the right-hand side (RHS) of 
Eq.~(\ref{gra}) as the {\em leading part (term)} and to the second 
term, which includes derivative $\partial_X$, as the {\em gradient} 
or {\em nonleading part (term)}. Crudely speaking, the leading term 
is the quantity in itself in the spacetime cell labeled by $X$, 
while the gradient term represents the effect arising from the 
(weak) correlation between the spacetime cell labeled by $(x + z) / 
2$ (see Eq.~(\ref{seki})) and the one labeled by $(z + y) / 2$.

The perturbative framework to be constructed accompanies the 
generalized Boltzmann equation (GBE) for the number density of 
quasiparticles. The framework allows us to compute any 
reaction rate by using the {\em reaction-rate formula} deduced in 
\cite{noo}. Substituting the computed net production rates of 
quasiparticles into the GBE, one can determine the number densities 
as functions of spacetime coordinates $X$, which describes the 
evolution of the system. 

For the sake of concrete presentation, we take up QCD. However, 
the procedure has little to do with QCD and then the framework to be 
constructed may be used for any theory that includes massless Dirac 
fermion(s) with almost no modification. The plan of the paper is as 
follows: In Sec.~II, making use of the free quark (fermion) fields 
in vacuum theory and the quark-distribution function at the initial 
time ($X^0 = - \infty$), we construct the bare quark propagator, 
with which the \lq\lq bare-$N$ scheme'' may be constructed. The 
perturbative calculation on the basis of this scheme yields 
\cite{nie} divergence due to pinch singularities. Then, in Sec.~III, 
we set up the basis for formulating the \lq\lq physical-$N$ scheme'' 
by introducing a new quark-distribution function and a new \lq\lq 
free'' quark fields. In Sec.~IV, we make up the self-energy-part 
resummed quark propagator. In Sec.~V, imposing the condition on the 
quark-distribution function that there do not appear large 
contributions, which stems from the above-mentioned pinch 
singularities, we construct a \lq\lq healthy'' perturbative 
framework. It is shown that, on the energy-shell, the condition 
turns out to be the generalized Boltzmann equation. In Sec.~VI, we 
frame a concrete perturbative framework. Section VII is devoted to 
summary and discussion, in which comparison with related works is 
made. Concrete derivation of various formula used in the text is 
made in Appendices. 
%%%%%%%%%%%%%%%%%%%%%%%%%%%%%%%%%%%%%%%%%%%%%%
%%%% SECTION I %%%%%%%%%%%%%%%%%%%%%%%%%%%%%%%
%%%%%%%%%%%%%%%%%%%%%%%%%%%%%%%%%%%%%%%%%%%%%%
\setcounter{equation}{0}
\setcounter{section}{1}
\def\theequation{\mbox{\arabic{section}.\arabic{equation}}} 
\section{Closed-time-path formalism} 
\subsection{Preliminary} 
We follow the procedure as in \cite{nie2}. The CTP formalism is 
reduced to a two-component formalism \cite{chou}. Every field, say 
$\phi$, is doubled, $\phi \to (\phi_1, \phi_2)$, and the classical 
action turns out 
%%%%%%%%%%%%%%%%%%%%%%%%%%%%%%%%%%%%%
\begin{equation} 
\int_{- \infty}^{+ \infty} d x_0 \int d {\bf x} \left[ {\cal L} 
(\phi_1 (x), ...) - {\cal L} (\phi_2 (x), ...) \right]  
\label{ha} 
\end{equation} 
%%%%%%%%%%%%%%%%%%%%%%%%%%%%%%%%%%%%%%%%%%%%%
with ${\cal L}$ the Lagrangian density of the theory under 
consideration. For definiteness, we take up massless QCD. The gluon 
sector has already been studied in \cite{nie2}. In this paper, we 
deal with the quark sector, 
%%%%%%%%%%%%%%%%%%%%%%%%%%%%%%%%%%%%%%%%%%%%%%%%%%%%%%%%%%%%%%%%
\begin{equation} 
{\cal L} = i \bar{\psi}^a (x) 
{\partial\mbox{\hspace*{0.3mm}}\kern-0.1em\raise0.3ex\llap{/}
\kern0.15em\relax} \psi^a (x) + ... , 
\label{ni} 
\end{equation} 
%%%%%%%%%%%%%%%%%%%%%%%%%%%%%%%%%%%%%%%%%%%%%%%%%%%%%%%%%%%%%%%%
where \lq\lq $a$'' is a color index. 
Perturbation theory is formulated in terms of propagators, vertices, 
and initial correlations\footnote{Given a density matrix, which 
characterize the system, the initial correlations are determined 
\cite{chou,lm}. In perturbation theory, the initial correlations are 
treated as vertexes.}. The aim of the present paper is to construct 
the quark propagator, which takes the $2 \times 2$ matrix form 
\cite{chou}, 
\widetext 
%%%%%%%%%%%%%%%%%%%%%%%%%
\begin{equation} 
\hat{S}^{a b}_{\alpha \beta} (x, y) \equiv \left( 
\begin{array}{cc}
- i \langle T (\psi_1^a (x))_\alpha (\bar{\psi}_1^b 
(y))_\beta \rangle & \;\; 
+ i \langle (\bar{\psi}_2^b (y))_\beta (\psi_1^a (x))_\alpha \, 
\rangle \\ 
- i \langle (\psi_2^a (x))_\alpha (\bar{\psi}_1^b (y))_\beta \, 
\rangle & \;\; 
- i \langle \bar{T} (\psi_2^a (x))_\alpha 
(\bar{\psi}_2^b (y))_\beta \rangle 
\end{array}
\right) . 
\label{matrix} 
\end{equation} 
%%%%%%%%%%%%%%%%%%%%%%%%%%%%%%%%%%%%%%%%%%%%%%
\narrowtext 

\noindent Here $\langle ... \rangle \equiv \mbox{Tr} \left( ... \, 
\rho \right)$ with $\rho$ the density matrix at the initial time, 
the \lq caret' denotes the $2 \times 2$ matrix as indicated on the 
RHS, the suffix $\alpha$ $(\beta)$ denotes the component of 
$\psi_j^a$ ($\bar{\psi}_j^a$) [$j = 1, 2$], and $T$ ($\bar{T}$) is 
the time-ordering (antitime-ordering) symbol. Noting that the state 
realized at a heavy-ion collision is color singlet, we restrict 
ourselves to the case where $\rho$ is color singlet, so that 
$\hat{S}^{a b}_{\alpha \beta} = \delta^{a b} \hat{S}_{\alpha 
\beta}$. We also assume that $\rho$ commutes with baryonic charge. 
[Generalization to the color-nonsinglet case is straightforward.] In 
the sequel, we drop the color index. At the end of calculation we 
set $\psi_1 = \psi_2$ and $\bar{\psi}_1 = \bar{\psi}_2$ \cite{chou}. 
Instructive examples of $\rho$, in the case of scalar field, is 
given in \cite{ch}. 
%%%%%%%%%%%%%%%%%%%%%%%%%%%%%%%%%%%%%%%%%%%%%%%%%%%%
\subsection{Quark propagator} 
We start with computing Eq.~(\ref{matrix}) with $\psi_1 = \psi_2$ 
$(\equiv \psi)$ and $\bar{\psi}_1 = \bar{\psi}_2$ $(\equiv 
\bar{\psi})$. For $\psi$ and $\bar{\psi}$, we use the 
plane-wave decomposition with helicity basis in vacuum theory, 
which, in standard notation, reads 
%%%%%%%%%%%%%%%%%%%%%%%%%
\begin{eqnarray} 
\psi (x) & = & \int \frac{d^{\, 3} p}{\sqrt{(2 \pi)^3 2 p}} 
\sum_{\xi = \pm} \left[ b_\xi (p) u_\xi (p) e^{- i P \cdot x} 
\right. \nonumber \\ 
&& \left. \mbox{\hspace*{3ex}} + d^\dagger_\xi (p) v_\xi (p) e^{i P 
\cdot x} \right] \mbox{\hspace*{8ex}} (p^0 = p) , 
\label{ten} 
\end{eqnarray} 
%%%%%%%%%%%%%%%%%%%%%%%%%%%%%%%%%%%%%%%%%%%%%%
where $\xi$ ($= \pm$) denotes helicity, and $b_\xi (p)$ 
[$b^\dagger_\xi (p)$] is the annihilation [creation] operator of a 
quark [an antiquark] with helicity $\xi$. Substituting 
Eq.~(\ref{ten}) and its adjoint into Eq.~(\ref{matrix}), we compute 
the Wigner transform, $\hat{S} (X; P)$, of $\hat{S} (x, y)$ [cf. 
Eq.~(\ref{Fou})] to the gradient approximation. There emerge four 
types of terms, which contain $\langle b^\dagger_\xi b_\zeta 
\rangle$, $\langle d^\dagger_\xi d_\zeta \rangle$, $\langle b_\xi 
d_\zeta \rangle$, and $\langle b^\dagger_\xi d^\dagger_\zeta 
\rangle$, respectively. For an equilibrium system, only $\langle 
b^\dagger_\pm b_\pm \rangle$ and $\langle d^\dagger_\pm d_\pm 
\rangle$ terms survive, and $\langle b^\dagger_+ b_+ \rangle = 
\langle b^\dagger_- b_- \rangle$ and $\langle d^\dagger_+ d_+ 
\rangle = \langle d^\dagger_- d_- \rangle$ hold. With this 
observation in mind, we see that, for the systems of our interest, 
the leading term of $\hat{S} (X; P)$, which we write $\hat{S}_0 (X; 
P)$, contains the \lq\lq even'' combinations $\sum_{\xi = \pm} 
\langle b^\dagger_\xi b_\xi \rangle$ and $\sum_{\xi = \pm} \langle 
d^\dagger_\xi d_\xi \rangle$. It turns out that the gradient term 
$\hat{S}_1 (X; P)$ contains also $\sum_{\xi = \pm} \langle 
b^\dagger_\xi b_\xi \rangle$ and $\sum_{\xi = \pm} \langle 
d^\dagger_\xi d_\xi \rangle$ together with spacetime ($X$) 
derivative. We assume that the term $\hat{S}_2 (X; P)$ that contains 
the \lq\lq odd'' combinations $\sum_{\xi = \pm} \xi \langle 
b^\dagger_\xi b_\xi \rangle$ and $\sum_{\xi = \pm} \xi \langle 
d^\dagger_\xi d_\xi \rangle$ and the term $\hat{S}_3 (X; P)$ that 
contains $\langle b^\dagger_\pm b_\mp \rangle$ and $\langle 
d^\dagger_\pm d_\mp \rangle$ are nonleading and are, at most, of the 
same order of magnitude as the gradient term $\hat{S}_1 (X; P)$. We 
also assume that the terms with $\langle b_\xi d_\zeta \rangle$ and 
with $\langle b^\dagger_\xi d^\dagger_\zeta \rangle$ may be ignored. 
It is a straightforward task to obtain the form of $\hat{S} (X; P)$ 
for the case where some or all of the above assumptions do not hold. 

Thus, $\hat{S} (X; P)$ consists of four terms, the leading term 
$\hat{S}_0$, the gradient term $\hat{S}_1$, and the nonleading terms 
$\hat{S}_2 + \hat{S}_3$, 
\widetext 
%%%%%%%%%%%%%%%%%%%%%%%%%
\begin{equation} 
\hat{S} (X; P) = \hat{S}_0 (X; P) + \hat{S}_1 (X; P) + \hat{S}_2 
(X; P) + \hat{S}_3 (X; P) . 
\label{bun} 
\end{equation}  
%%%%%%%%%%%%%%%%%%%%%%%%%%%
The form for $\hat{S}_2$ and $\hat{S}_3$ are displayed in Appendix 
A. The terms $\hat{S}_0$ and $\hat{S}_1$ reads 
%%%%%%%%%%%%%%%%%%%%%%%%%
\begin{eqnarray} 
\hat{S}_0 (X; P) & = & 
{P\kern-0.1em\raise0.3ex\llap{/}\kern0.15em\relax} 
\left( 
\begin{array}{cc} 
\Delta_R (P) & \;\; 0 \\ 
\Delta_R (P) - \Delta_A (P) & \;\; - \Delta_A (P) 
\end{array} 
\right) - f (X; P) 
{P\kern-0.1em\raise0.3ex\llap{/}\kern0.15em\relax}  
\left[ \Delta_R (P) - \Delta_A (P) \right] \hat{A}_+ , 
\label{s0} \\ 
\hat{S}_1 (X; P) & = & - \frac{\epsilon^{0 \mu \nu \rho} + 
\epsilon^{3 \mu \nu \rho}}{2 (p^0 + p^3)} P_\mu \gamma_5 \gamma_\rho 
\frac{\partial f (X; P)}{\partial X^\nu} \left[ \Delta_R (P) - 
\Delta_A (P) \right] \hat{A}_+ . 
\label{s1} 
\end{eqnarray} 
%%%%%%%%%%%%%%%%%%%%%%%%%%%%%%%%%%%%%%%%%%%%%%
Here $\epsilon^{0 1 2 3} = 1$, $\gamma_5 = i \gamma^0 \gamma^1 
\gamma^2 \gamma^3$, and 
%%%%%%%%%%%%%%%%%%%%%%%%%%%%%%%%%%%%%%%%%%%%%%%%%%%%%%%%%%%%%
\begin{eqnarray} 
\hat{A}_{\pm} & = & \left( 
\begin{array}{cc} 
1 & \;\; \pm1 \\ 
\pm 1 & 1 
\end{array} 
\right) , 
\label{Apm} \\ 
\Delta_{R (A)} (P) & = & \frac{1}{P^2 \pm i p_0 0^+} , 
\label{RA} \nonumber \\ 
f (X; P) & = & \theta (- p_0) + \epsilon (p_0) N (X; P) , 
\label{f-desu} 
\\ 
N (X; P) & \equiv & \frac{1}{2} \left[ \theta (p^0) 
\left\{ N_{+ +} (X; |p^0|, \tilde{\bf p}) +  N_{- -} (X; |p^0|, 
\tilde{\bf p}) \right\} \right. \nonumber \\ 
&& \left. + \theta (- p^0) \left\{ \bar{N}_{+ +} (X; |p^0|, - 
\tilde{\bf p}) + \bar{N}_{- -} (X; |p^0|, - \tilde{\bf p}) \right\} 
\right] \nonumber \\ 
& \equiv & \theta (p^0) n (X; |p^0|, \tilde{\bf p}) + \theta (- p^0) 
\bar{n} (X; |p^0|, - \tilde{\bf p}) , 
\label{N-desu} 
\end{eqnarray} 
%%%%%%%%%%%%%%%%%%%%%%%%%%%%%%%%%%%%%%%%%%%%%%
with 
%%%%%%%%%%%%%%%%%%%%%%%%%%%%%%%%%%%%%%%%%%%%%%%%%%%%%%%%%%%%%
\begin{eqnarray} 
N_{\xi \zeta} (X; |p_0|, \tilde{\bf p}) & \equiv & \int d^{\, 3} q 
\, e^{- i Q \cdot X} \mbox{Tr} \left[ 
b^\dagger_\xi ({\bf p} - {\bf q} / 2) \, b_\zeta ({\bf p} + {\bf 
q} / 2) \, \rho \right] 
\;\;\; (\xi, \zeta = +, -) . 
\label{no} 
\end{eqnarray} 
%%%%%%%%%%%%%%%%%%%%%%%%%%%%%%%%%%%%%%%%%%%%%%
\narrowtext 

\noindent Here $|p_0| = p$, $\tilde{\bf p} \equiv {\bf p} / p$, 
and $Q = (q^0, {\bf q})$ with $q_0 = {\bf q} \cdot {\bf p} / 
p$. $\bar{N}$'s are antiquark counterpart of $N$'s. For the system of 
our concern, $\mbox{Tr} [...]$ is different from from zero 
\cite{chou,lm} only for $q$ small compared with ${\bf p}$. 
It can readily be seen from Eq.~(\ref{no}) that when $\rho$ is 
translationally invariant $N_{\xi \zeta} (X; |p^0|, \tilde{\bf p})$ 
is independent of $X$. In passing, for a 
charge-conjugation-invariant system, $N (X; P) = N (X; - P)$ holds. 

From Eq.~(\ref{no}) follows 
%%%%%%%%%%%%%%%%%%%%%%%
\begin{eqnarray} 
&& \theta (p^0) P \cdot \partial_X N_{\xi \zeta} (X; |p^0|, 
\tilde{\bf p}) \nonumber \\ 
&& \mbox{\hspace*{5ex}} = \theta (- p^0) P \cdot \partial_X 
\bar{N}_{\xi \zeta} (X; |p^0|, - \tilde{\bf p}) = 0 
\;\;\;\;\;\; (|p_0| = p) . \nonumber \\ 
\label{ba} 
\end{eqnarray} 
%%%%%%%%%%%%%%%%%%%%%%%%%%
Using this in Eqs.~(\ref{f-desu}) and (\ref{N-desu}), we 
have 
%%%%%%%%%%%%%%%%%%%%%%%
\begin{equation} 
P \cdot \partial_X f (X; P) = 0 \;\;\;\;\;\; (|p_0| = p) . 
\label{ba1} 
\end{equation} 
%%%%%%%%%%%%%%%%%%%%%%%%%%
As is obvious from the construction, or as can be directly shown, 
to the gradient approximation, 
%%%%%%%%%%%%%%%%%%%%%%%%%%%%%%%%%%%%%%%%%%%%%%%%%%%%%%%%%%%%%%%
\[ 
i \hat{\tau}_3 
{\partial\mbox{\hspace*{0.3mm}}\kern-0.1em\raise0.3ex\llap{/}
\kern0.15em\relax}_x \hat{S} 
(x, y) = - i \hat{S} (x, y) 
\stackrel{\leftarrow}{
\partial\mbox{\hspace*{0.3mm}}\kern-0.1em\raise0.3ex\llap{/}
\kern0.15em\relax}_y \hat{\tau}_3 = \hat{1} \, \delta^{\, 4} (x - y) 
, 
% \label{aa} 
\] 
%%%%%%%%%%%%%%%%%%%%%%%%%%%%%%%%%%%%%%%%%%%%%%%%
where $\hat{1}$ is the $2 \times 2$ unit matrix and $\hat{\tau}_3$ 
is the third Pauli matrix. Two equations in 
Eq.~(\ref{ba}) are \lq\lq free Boltzmann equations.'' One can 
construct a perturbation theory in a similar manner as in 
\cite{nie}. We call the perturbation theory thus constructed the 
bare-$N$ scheme, since $N$ obeys the \lq\lq free Boltzmann 
equation.'' Perturbative computation within this scheme yields 
divergences due to pinch singularities. In \cite{nie}, how to deal 
with these divergences is discussed and shown is that the bare-$N$ 
scheme is equivalent to the physical-$N$ scheme, to which we now 
turn. 
%%%%%%%%%%%%%%%%%%%%%%%%%%%%%%%%%%%%%%%%%%%%%%
%%%% SECTION III %%%%%%%%%%%%%%%%%%%%%%%%%%%%%%%
%%%%%%%%%%%%%%%%%%%%%%%%%%%%%%%%%%%%%%%%%%%%%%
%\narrowtext 
\setcounter{equation}{0}
\setcounter{section}{2}
\def\theequation{\mbox{\arabic{section}.\arabic{equation}}} 
\section{Construction of the physical-$N$ scheme} 
Following the procedure as in \cite{nie2}, we construct a scheme in 
terms of the number density that is as close as possible to the 
physical number density. To this end, first of all, we abandon the 
\lq\lq free Boltzmann equation'' (\ref{ba1}). This means that $f$ 
here and in the sequel is different from $f$ ($\equiv f_B$) in the 
last section. Then $\hat{S} (X; P; f) \neq \hat{S} (X; P; f_B)$, 
which means that $\psi$, and $\bar{\psi}$ in the present 
(physical-$N$) scheme differs from the free field in vacuum theory, 
Eq.~(\ref{ten}), as employed in the bare-$N$ scheme. [See also the 
comment below in conjunction with Eq.~(\ref{acti}).] Specification 
of $f$ and then also of $\psi$ and $\bar{\psi}$ are postponed until 
Sec.~VI. 

Now, $\hat{S} (x, y)$ is not an inverse of $i \hat{\tau}_3 
{\partial\mbox{\hspace*{0.3mm}}\kern-0.1em\raise0.3ex\llap{/}
\kern0.15em\relax}$ [cf. Eq.~(\ref{ha}) with Eq.~(\ref{ni})]. 
Straightforward computation using the inverse Wigner transform of 
$\hat{S} (X; P)$, Eq.~(\ref{bun}), yields, to the gradient 
approximation, 
%%%%%%%%%%%%%%%%%%%%%%%%%%%%%%%%
\begin{eqnarray} 
i \hat{\tau}_3 
{\partial\mbox{\hspace*{0.3mm}}\kern-0.1em\raise0.3ex\llap{/}
\kern0.15em\relax}_x 
\hat{S} (x, y) & = & \hat{1} \, \delta^{\, 4} (x - y) - \frac{i}{2} 
\hat{\tau}_3 \hat{A}_+ \int \frac{d^{\, 4} P}{(2 \pi)^4} e^{- i P 
\cdot (x - y)} \nonumber \\ 
&& \times \frac{\gamma^0 + \gamma^3}{p^0 + p^3} 
{P\kern-0.1em\raise0.3ex\llap{/}\kern0.15em\relax}  
(\Delta_R - \Delta_A) P \cdot 
\partial_X f (X; P) . \nonumber \\ 
\label{shu} 
\end{eqnarray} 
%%%%%%%%%%%%%%%%%%%%%%%%%%%%%%%%%%%%%%%%%%%%%
Our procedure of constructing a consistent scheme is as follows: 
We further modify $\psi$ and $\bar{\psi}$ by adding a suitable 
$\hat{S}_{\mbox{\scriptsize{add}}}$ to $\hat{S}$ on the left-hand 
side (LHS) of Eq.~(\ref{shu}). The conditions for 
$\hat{S}_{\mbox{\scriptsize{add}}}$ to be satisfied are 
\begin{itemize} 
\item $\hat{S}_{\mbox{\scriptsize{add}}}$ vanishes in the bare-$N$ 
scheme in Sec.~II. 
\item To the gradient approximation, Eq.~(\ref{shu}) turns out to 
%%%%%%%%%%%%%%%%%%%%%%%%%%%%%%%%%%%%%%%%%%%%%
\begin{equation} 
\left( i {\hat{\tau}}_3 
{\partial\mbox{\hspace*{0.3mm}}\kern-0.1em\raise0.3ex\llap{/}
\kern0.15em\relax} 1 - \hat{L}_c \right) \cdot \left( \hat{S} + 
\hat{S}_{\mbox{\scriptsize{add}}} \right) = \hat{1} \, 1 . 
\label{ma} 
\end{equation} 
%%%%%%%%%%%%%%%%%%%%%%%%%%%%%%%%
Here we have used the short-hand notation (\ref{seki}). In 
Eq.~(\ref{ma}), \lq $1$' is the matrix in spacetime whose $(x, 
y)$-component is $\delta^{\, 4} (x - y)$, and $\hat{L_c}$ is some 
$(4 \times 4) \otimes (2 \times 2)$ matrix function. 
\end{itemize} 

It is straightforward to obtain the form of the required 
$\hat{S}_{\mbox{\scriptsize{add}}}$: 
%%%%%%%%%%%%%%%%%%%%%%%%%%%%%%%%
\begin{eqnarray} 
\hat{S}_{\mbox{\scriptsize{add}}} (X;. P) & = & i \hat{A}_+ \left[ 
\left( \Delta_R^2 + \Delta_A^2 \right) 
{P\kern-0.1em\raise0.3ex\llap{/}\kern0.15em\relax} - 
\frac{\gamma^0 + \gamma^3}{p^0 + p^3} \frac{\bf P}{P^2} \right] 
\nonumber \\ 
&& \times P \cdot \partial_X f (X; P) , 
\label{add} 
\end{eqnarray} 
%%%%%%%%%%%%%%%%%%%%%%%%%%%%%%%%%%%%%%%%%%%%%
(${\bf P} / P^2$ the principal part of $1 / (P^2 \pm i 0^+)$) from 
which we obtain for $\hat{L}_c$ in Eq.~(\ref{ma}), 
%%%%%%%%%%%%%%%%%%%%%%%%%%%%%%%%
\begin{eqnarray*} 
\hat{L}_c (x, y) & = & L_c \hat{A}_- = i \hat{A}_- \int \frac{d^{\, 
4} P}{(2 \pi)^4} e^{- i P \cdot (x - y)} \frac{\gamma^0 + 
\gamma^3}{p^0 + p^3} \nonumber \\ 
&& \mbox{\hspace*{15ex}} \times P \cdot \partial_X f (X; P) , 
\end{eqnarray*} 
%%%%%%%%%%%%%%%%%%%%%%%%%%%%%%%%%%%%%%%%%%%%%
with $\hat{A}_-$ as in Eq.~(\ref{Apm}). In obtaining Eq.~(\ref{ma}) 
with $\hat{S}_{\mbox{\scriptsize{add}}}$ as in Eq.~(\ref{add}), we 
have used $\hat{L}_c \cdot \hat{S}_0 \simeq \hat{L}_c \cdot (\hat{S} 
+ \hat{S}_{\mbox{\scriptsize{add}}} )$, since the difference can be 
ignored to the gradient approximation.  

In a similar manner, we find that $(\hat{S} + 
\hat{S}_{\mbox{\scriptsize{add}}}) \cdot (- i \hat{\tau}_3 1 
\stackrel{\leftarrow}{{\partial\mbox{\hspace*{0.3mm}}
\kern-0.1em\raise0.3ex\llap{/}\kern0.15em\relax}} - \hat{L}_c) = 
\hat{1} \, 1$. Thus we have found that the $(2 \times 2) \otimes (4 
\times 4)$ matrix propagator $(\hat{S} + 
\hat{S}_{\mbox{\scriptsize{add}}})$ is an inverse of $(i 
\hat{\tau}_3 {\partial\mbox{\hspace*{0.3mm}}
\kern-0.1em\raise0.3ex\llap{/}\kern0.15em\relax} 1 - \hat{L}_c)$, so 
that the free action is
%%%%%%%%%%%%%%%%%%%%%%%%%%%%%%%%%%%%%%%%%%%%%%%%%%%%%%%%%%%%%
\begin{eqnarray} 
&& \int d^{\, 4} x \, d^{\, 4} y \, \tilde{\bar{\psi}} (x) \left[ 
i \hat{\tau}_3 
{\partial\kern-0.1em\raise0.3ex\llap{/}\kern0.15em\relax}_x 
\delta^{\, 4} (x - y) - L_c (x, y) \hat{A}_- \right] \tilde{\psi} 
(y) , \nonumber \\ 
&& \mbox{\hspace*{5ex}} 
\tilde{\bar{\psi}} = \left( \bar{\psi}_1, \, \bar{\psi}_2 \right) 
\, , \;\;\;\;\;\;\;\;\; \tilde{\psi} = \left( 
\begin{array}{c} 
\psi_1 \\ 
\psi_2 
\end{array} 
\right) . 
\label{acti} 
\end{eqnarray} 
%%%%%%%%%%%%%%%%%%%%%%%%%%%%%%%%%%%%%%%%%%%%%%%%%%%%%%%
This is different from the free action in the bare-$N$ scheme, 
$\int d^{\, 4} x \, \tilde{\bar{\psi}} i \hat{\tau}_3 
{\partial\kern-0.1em\raise0.3ex\llap{/}\kern0.15em\relax} 
\tilde{\psi}$, and then the free fields in Eq. (\ref{acti}) is 
different from the free fields employed in the bare-$N$ scheme. 
Since the term with $L_c (x, y)$ in Eq.~(\ref{acti}) is absent in 
the original action, we should introduce the counter action to 
compensate it, 
%%%%%%%%%%%%%%%%%%%%%%%%%%%%%%%%%%%%%%%%%%%%
\begin{equation} 
{\cal A}_c = \int d^{\, 4} x \, d^{\, 4} y \, \tilde{\bar{\psi}} (x) 
L_c (x, y) \hat{A}_- \tilde{\psi} (y) , 
\label{act} 
\label{coun} 
\end{equation} 
%%%%%%%%%%%%%%%%%%%%
which yields a vertex 
%%%%%%%%%%%%%%%%%%%%%%%%%%%%%%%%%%%%%%%%%%%%
\begin{eqnarray} 
i L_c (x, y) \hat{A}_- & = & - \hat{A}_- \int \frac{d^{\, 4} P}{(2 
\pi)^4} e^{- i P \cdot (x - y)} \frac{\gamma^0 + \gamma^3}{p^0 + 
p^3} \nonumber \\ 
&& \times P \cdot \partial_X f (X; P) \nonumber \\ 
& \equiv & - i \hat{\Sigma}^{(c)} (x, y) . 
\label{sigmac} 
\end{eqnarray} 
%%%%%%%%%%%%%%%%%%%%

As in \cite{nie2}, it is straightforward task to construct a 
$\hat{L}_c$-resummed propagator, which we do not reproduce here. 

In closing this section, we emphasize that $f (X; P)$ in the present 
scheme is an arbitrary function, provided that $f (X^0_{in}, {\bf 
X}; P)$ [$= \theta (- p^0) + \epsilon (p^0) N(X^0_{in}, {\bf X}; 
P)$], with $X^0_{in}$ $(= - \infty)$ the initial time, is a given 
initial data. We have introduced the counteraction ${\cal A}_c$, 
Eq.~(\ref{coun}), so as to remain on the original theory. Thus, it 
cannot be overemphasized that the schemes with different $f$'s are 
mutually equivalent. If we choose $f_B$ for $f$, which subjects to 
the \lq\lq free Boltzmann equation'' (\ref{ba1}), the scheme reduces 
to the bare-$N$ scheme in the last section. In Sec.~VI, we shall 
choose $f$, with which a well-defined perturbation theory is 
formulated. As will be shown below, it turns out that $f$ is 
determined order by order in perturbation theory. As a natural 
assumption, we require that, in the limit $g \to 0$ ($g$ the QCD 
coupling constant), $f \to f_B$. 
%%%%%%%%%%%%%%%%%%%%%%%%%%%%%%%%%%%%%%%%%%%%%%
%%%% SECTION III %%%%%%%%%%%%%%%%%%%%%%%%%%%%%%%
%%%%%%%%%%%%%%%%%%%%%%%%%%%%%%%%%%%%%%%%%%%%%%
%\narrowtext 
\setcounter{equation}{0}
\setcounter{section}{3}
\def\theequation{\mbox{\arabic{section}.\arabic{equation}}} 
\section{Resummation of the self-energy part} 
%%%%%%%%%%%%%%%%%%%%%%%%%%%%%%%%%%%%%%%%%%%%
\subsection{Preliminary} 
As has been mentioned at the end of last section, when interactions 
are switched off, the self-energy part $\hat{\Sigma}$ vanishes and 
Eq.~(\ref{ba1}) holds, $P \cdot \partial_X f (X; P) = 0$. Thus, we 
suspect that the self-energy part causes the nontrivial evolution of 
$f$, $P \cdot \partial_X f (X; P) \neq 0$. Thus, $\hat{\Sigma}$ ties 
to the gradient part, $\hat{S}_1$, of $\hat{S}$. More precisely, 
$\hat{\Sigma}$ is of the same order of magnitude as $(\hat{S}_0)^{- 
1} \hat{S}_1 (\hat{S}_0)^{- 1}$ and $(\hat{S}_0)^{- 1} 
\hat{S}_{\mbox{\scriptsize{add}}} (\hat{S}_0)^{- 1}$. Hence, in 
computing $\hat{\Sigma}$ in the approximation under consideration, 
it is sufficient to keep the leading part (i.e., the part with no 
$X_\mu$-derivative). 

Since we are dealing with the massless quark, the leading part 
$\hat{\Sigma} (X; P)$ may be decomposed as $\hat{\Sigma} = [ 
\tilde{\sigma}^+ 
{\tilde{P}\kern-0.1em\raise0.3ex\llap{/}\kern0.15em\relax}_- - 
\tilde{\sigma}^- 
{\tilde{P}\kern-0.1em\raise0.3ex\llap{/}\kern0.15em\relax}_+ ] / 2 + 
\sum_{a = 1}^2 \hat{\sigma}_a \vec{\gamma} \cdot \vec{e}_a (p)$, 
where 
%%%%%%%%%%%%%%%%%%%%%%%%%%%%%%%%%%%%%%%%%%%%
\begin{equation} 
\tilde{P}_\pm \equiv \left(1, \, \pm \tilde{\bf p} \right) \;\;\;\;\;
\; (\tilde{\bf p} \equiv {\bf p} / p) , 
\label{ppm} 
\end{equation} 
%%%%%%%%%%%%%%%%%%%%%%%%%%%%%%%%%%%%%%%%%%%%
and $\vec{e}_1 (p)$ and $\vec{e}_2 (p)$ are unit vectors being 
orthogonal to ${\bf p}$; ${\bf p} \cdot \vec{e}_a (p) = \vec{e}_1 
(p) \cdot \vec{e}_2 (p) = 0$ ($a = 1, 2$). For an equilibrium 
system, $\hat{\sigma_1} = \hat{\sigma_2} = 0$. Then it is natural to 
assume that, for the systems of our concern, $\hat{\sigma}_a$ $(a = 
1, 2)$ can be ignored when compared to the leading part, 
%%%%%%%%%%%%%%%%%%%%%%%%%%%%%%%%%%%%
\[ 
\hat{\Sigma} (X; P) = \frac{1}{2} \left[ \hat{\sigma}^+ (X; P) 
{\tilde{P}\kern-0.1em\raise0.3ex\llap{/}\kern0.15em\relax}_-  
- \hat{\sigma}^- (X; P) 
{\tilde{P}\kern-0.1em\raise0.3ex\llap{/}\kern0.15em\relax}_+ \right] 
. 
% \label{koo} 
\] 
%%%%%%%%%%%%%%%%%%%%%%%%%%%%%%%%%%%%%%%%%%%%
Generalization to the case where this assumption does not hold is 
straightforward. It should be noted that $\hat{\Sigma}$ consists of 
two pieces, 
%%%%%%%%%%%%%%%%%%%%%%%%%%%%%%%%%%
\[ 
\hat{\Sigma} = \hat{\Sigma}^{\mbox{\scriptsize{loop}}} + 
\hat{\Sigma}^{(c)} , 
% \label{two} 
\] 
%%%%%%%%%%%%%%%%%%%%%%%%%%%%%%%%%%
where $\hat{\Sigma}^{\mbox{\scriptsize{loop}}}$ is the contribution 
from loop diagrams and $\hat{\Sigma}^{(c)}$ is as in 
Eq.~(\ref{sigmac}). It should be remarked that some 
$\hat{\Sigma}^{\mbox{\scriptsize{loop}}}$ contains internal 
vertex(es) $i \hat{\Sigma}^{(c)}$. [See Sec.~VI 
below.] 

Within the gradient approximation, it is sufficient to perform a 
$\hat{\Sigma}$-resummation for the leading part $\hat{S}_0$. This 
is because the corrections to other parts due to the resummation 
are of higher order. Thus, for $\hat{S}_1 \, - \, \hat{S}_3$ and 
$\hat{S}_{\mbox{\scriptsize{add}}}$, one can use the formulae in 
the bare-$N$ scheme in Sec.~II [see the argument at the end of 
Sec.~III]. In particular, for $f$ in $\hat{S}_1 \, - \, \hat{S}_3$ 
and $\hat{S}_{\mbox{\scriptsize{add}}}$, one can use $f_B$. Now 
we introduce, as usual, the \lq\lq standard form'' (see 
Eq.~(\ref{seki})) \cite{ume,nie2} 
%%%%%%%%%%%%%%%%%%%%%%%%%%%%%%%%%%%%%%%%%%%%
\begin{eqnarray} 
&& \left[ 
\hat{B}_L \cdot \hat{S}_{\mbox{\scriptsize{diag}}} \cdot \hat{B}_R 
\right] (x, y) , 
\label{standard} \\ 
&& \hat{S}_{\mbox{\scriptsize{diag}}} = \mbox{diag} ( i 
{\partial\kern-0.1em\raise0.3ex\llap{/}\kern0.15em\relax} \Delta_R, 
\, - i {\partial\kern-0.1em\raise0.3ex\llap{/}\kern0.15em\relax} 
\Delta_A) , \nonumber \\ 
&& \hat{B}_L = \left( 
\begin{array}{cc} 
1 & \;\; - f \\ 
1 & \;\; 1 - f 
\end{array} 
\right) \, , \;\;\;\;\; 
\hat{B}_R = \left( 
\begin{array}{cc} 
1 - f & \;\; - f \\ 
1 & \;\; 1 
\end{array} 
\right) , 
\label{4.111} 
\end{eqnarray} 
%%%%%%%%%%%%%%%%%%%%
where $f$ $(= f (x, y))$ is the inverse Wigner transform of $f (X; 
P)$ and $1 (x, y) = \delta^{\, 4} (x - y)$. Computing 
Eq.~(\ref{standard}) to the gradient approximation, we obtain 
%%%%%%%%%%%%%%%%%%%%%%%%%%%%%%%%%%%%%%%%%%%%
\begin{eqnarray} 
\hat{\cal S} (x, y) & \equiv & \hat{S}_0 (x, y) + 
\hat{S}_{\mbox{\scriptsize{add}}} (x, y) \nonumber \\ 
& = & \left[ \hat{B}_L \cdot \hat{S}_{\mbox{\scriptsize{diag}}} 
\cdot \hat{B}_R \right] 
(x, y) + i \int \frac{d^{\, 4} P}{(2 \pi)^4} e^{- i P 
\cdot (x - y)} \nonumber \\ 
&& \times \left[ 
{\partial\mbox{\hspace*{0.3mm}}\kern-0.1em\raise0.3ex\llap{/}
\kern0.15em\relax}_X f (X; P) - \frac{\gamma^0 + \gamma^3}{p^0 + 
p^3} P \cdot \partial_X f (X; P) \right] \nonumber \\ 
&& \times \frac{\bf P}{P^2} \hat{A}_+ . 
\label{calS} 
\end{eqnarray} 
%%%%%%%%%%%%%%%%%%%%%%%%%%%%%%%%%%%%%%%%%%%%%%%
It is obvious that one can freely include gradient part(s) into the 
\lq\lq resummed part.'' For convenience, we include the gradient 
part $\hat{S}_{\mbox{\scriptsize{add}}}$, Eq.~(\ref{add}), and take 
$\hat{\cal{S}}$ in Eq.~(\ref{calS}) as the \lq\lq resummed part.'' 

It is to be noted that, from Eqs.~(\ref{s0}) and (\ref{add}) 
follows\footnote{Here and in the following, $(i j)$ element of a $(2 
\times 2)$ matrix $\hat{M}$ is denoted by $M_{i j}$.}  
%%%%%%%%%%%%%%%%%%%%%%%%%%%%%%%%%%%%%%%%%%%%
\begin{equation} 
\sum_{i, \, j = 1}^2 (-)^{i + j} {\cal S}_{i j} = 
\sum_{i, \, j = 1}^2 (-)^{i + j} \left[S_{i j} + 
\left( S_{\mbox{\scriptsize{add}}} \right)_{i j} \right] = 0 . 
\label{meme} 
\end{equation} 
%%%%%%%%%%%%%%%%%%%%%%%%%%%%%%%%%%%%%%%%%%%%
%%%%%%%%%%%%%%%%%%%%%%%%%%%%%%%%%%%%%%%%%%%%%%%
%%%%%% SUB SEC %%%%%%%%%%%%%%%%%%%%%%%%%%%%%%%%%%%%%%%%%%%%%%%%%%
%%%%%%%%%%%%%%%%%%%%%%%%%%%%%%%%%%%%%%
\subsection{Self-energy-part resummed propagator} 
A $\hat{\Sigma}$-resummed propagator $\hat{G}$ obeys the 
Sch\-win\-ger-Dyson (SD) equation: 
%%%%%%%%%%%%%%%%%%%%%%%%%%%%%%%%%%%%%%%%%%%%%%%%%%%%%%%%%%%%%%%%
\begin{equation} 
\hat{G} = \hat{\cal S} + \hat{\cal S} \cdot \hat{\Sigma} \cdot 
\hat{G} = \hat{\cal S} + \hat{G} \cdot \hat{\Sigma} \cdot 
\hat{\cal S} . 
\label{SD} 
\end{equation} 
%%%%%%%%%%%%%%%%%%%%%%%%%%%%
We recall that $\hat{G}$ obeys \cite{chou} the same relation as 
Eq.~(\ref{meme}), 
$\sum_{i, \, j = 1}^2 (-)^{i + j} G_{i j} = 0$. Using this and 
Eq.~(\ref{meme}) in Eq.~(\ref{SD}), we obtain 
%%%%%%%%%%%%%%%%%%%%%%%%%%%%
\begin{equation} 
\sum_{i, \, j = 1}^2 \Sigma_{i j} = 0 . 
\label{4.66} 
\end{equation} 
%%%%%%%%%%%%%%%%%%%%%%%%%%%%
Procedure of solving Eq.~(\ref{SD}) is given in Appendix B. The 
result for $\hat{G} (X; P)$ takes the form 
%%%%%%%%%%%%%%%%%%%%%%%%%%%%%%%%%%%%%%%%%%%%%
\begin{eqnarray} 
&& \hat{G} (X; P) \nonumber \\ 
&& \mbox{\hspace*{3ex}} \simeq 
\left( 
\begin{array}{cc} 
G_R (X; P) & \;\;\, 0 \\ 
G_R (X; P) - G_A (X; P) & \;\;\, - G_A (X; P) 
\end{array} 
\right) \nonumber \\ 
&& \mbox{\hspace*{6ex}} - \hat{A}_+ \left[ G_R (X; P) - G_A (X; P) 
\right] f (X; P) \nonumber \\ 
&& \mbox{\hspace*{6ex}} + \hat{A}_+ G_K (X; P) . 
\label{3.111} 
\end{eqnarray} 
%%%%%%%%%%%%%%%%%%%%%%%%%%%%%%%%%%%%%%%%%%%%%%
$G_{R (A)}$ consists of the leading piece and the nonleading piece, 
%%%%%%%%%%%%%%%%%%%%%%
\begin{equation} 
G_{R (A)} (X; P) \simeq G_{R (A)}^{(0)} (X; P) + G_{R (A)}^{(1)} 
(X; P) , 
\label{GRkota} 
\end{equation} 
%%%%%%%%%%%%%%%%%%%%%%
The leading piece reads 
%%%%%%%%%%%%%%%%%%%%%%
\begin{eqnarray} 
G_{R (A)}^{(0)} (X; P) & = & \frac{1}{2} \left[ g_{R (A)}^+ (X; P) 
\tilde{P\kern-0.1em\raise0.3ex\llap{/}\kern0.15em\relax}_+ + 
g_{R (A)}^- (X; P) 
\tilde{P\kern-0.1em\raise0.3ex\llap{/}\kern0.15em\relax}_- 
\right] , \nonumber \\ 
\label{RA0} \\ 
g_{R (A)}^\pm & = & \frac{1}{p^0 \mp p \mp \sigma_{R (A)}^\pm} 
, 
\label{4.21} 
\end{eqnarray} 
%%%%%%%%%%%%%%%%%%%%%%
where $\tilde{P}_\pm$ is as in Eq.~(\ref{ppm}) and $\sigma_{R 
(A)}^\tau = \sigma_{1 1}^\tau + \sigma_{1 2 (2 
1)}^\tau$. Although the nonleading piece $G^{(1)}_{R (A)}$ can be 
ignored to the gradient approximation, we have displayed it in 
Appendix B. 

$G_K$ in Eq.~(\ref{3.111}) reads 
%%%%%%%%%%%%%%%%%%%%%%
\begin{eqnarray} 
G_K & \simeq & G_K^{(0)} + G_K^{(1)} , 
\label{GKkota} 
\\ 
G_K^{(0)} & = & - \frac{i}{2} \sum_{\tau = \pm} \left[ \left( 
\tilde{P}_\tau \cdot \partial_X f - i \tau \left( \sigma^\tau_K 
\right)^{\mbox{\scriptsize{loop}}} \right. \right. \nonumber \\ 
&& \mbox{\hspace*{10ex}} \left. \left. + \tau \left\{ \mbox{Re} 
\sigma^\tau_R , \;\, f \right\}_{\mbox{\scriptsize{P.B.}}} \right) 
g^\tau_R g^\tau_A 
\tilde{P\kern-0.1em\raise0.3ex\llap{/}\kern0.15em\relax}_\tau 
\right] , 
\label{GK0} 
\\ 
G_K^{(1)} & = & \frac{i}{4} \sum_{{\cal H} = R, \, A} \sum_{\tau = 
\pm} \left[ \left( \tilde{P}_\tau \cdot \partial_X f \right. 
\right. \nonumber 
\\ 
&& \mbox{\hspace*{15.5ex}} \left. \left. + \tau \left\{ 
\sigma_{\cal H}^\tau, \;\, f \right\}_{\mbox{\scriptsize{P.B.}}} 
\right) \left( g^\tau_{\cal H} \right)^2 \right] 
\tilde{P\kern-0.1em\raise0.3ex\llap{/}\kern0.15em\relax}_\tau 
\nonumber \\ 
&& - \frac{i}{16} \frac{1}{p} \sum_{\tau = \pm} 
\left[ \tau 
\tilde{P\kern-0.1em\raise0.3ex\llap{/}\kern0.15em\relax}_\tau \left( 
\gamma^i_\perp \partial_i f \right) 
\tilde{P\kern-0.1em\raise0.3ex\llap{/}\kern0.15em\relax}_{- \tau} 
\left( g^+_R  - g^+_A + g^-_R \right. \right. \nonumber \\ 
&& \left. \left. \mbox{\hspace*{11ex}} - g^-_A - 2 i \, 
g^\tau_R g^{- \tau}_A \mbox{Im} (\sigma^+_R - \sigma^-_R) 
\right) \right] , \nonumber \\ 
\label{GK2} 
\end{eqnarray} 
%%%%%%%%%%%%%%%%%%%%%%%%%%%%%%%%%%%%%%%%%%%%%%%%%%%%%%%%%%%%
where $\gamma^i_\perp \equiv \gamma^i - (\vec{\gamma} \cdot 
\tilde{\bf p}) \tilde{p}^i$ and 
%%%%%%%%%%%%%%%%%%%%%%%%%%%%%%%%%%
\begin{equation} 
\left( \sigma^\pm_K \right)^{\mbox{\scriptsize{loop}}} \equiv 
[1 - f] \left( \sigma_{1 2}^\pm \right)^{\mbox{\scriptsize{loop}}} 
+ f \left( \sigma_{2 1}^\pm \right)^{\mbox{\scriptsize{loop}}} . 
\label{3.188} 
\end{equation} 
%%%%%%%%%%%%%%%%%%%%%%%%%%%%%%%%%%%%%
In obtaining Eq.~(\ref{GK2}), use has been made of $\left( 
\sigma^\pm_A \right)^* = \sigma^\pm_R$, which can straightforwardly 
be proved \cite{chou}. 
%%%%%%%%%%%%%%%%%%%%%%%%%%%%%%%%%%%%%%%%%%%%%%
%%%% SECTION III %%%%%%%%%%%%%%%%%%%%%%%%%%%%%%%
%%%%%%%%%%%%%%%%%%%%%%%%%%%%%%%%%%%%%%%%%%%%%%
%\narrowtext 
\setcounter{equation}{0}
\setcounter{section}{4}
\def\theequation{\mbox{\arabic{section}.\arabic{equation}}} 
\section{Generalized Boltzmann equation} 
\subsection{Energy-shell and physical number densities} 
For later use, referring to Eq.~(\ref{RA0}) with Eq.~(\ref{4.21}), 
we define the energy-shell for \lq\lq normal modes'' through 
%%%%%%%%%%%%%%%%%%%%%%
\begin{eqnarray} 
&& \mbox{Re} \left[ g^\pm_R (X; P) \right]^{- 1}_{p^0 = \pm 
\omega_\pm^n (X; \pm {\bf p})} \nonumber \\ 
&& \mbox{\hspace*{5ex}} 
= \left[ p^0 \mp p \mp \mbox{Re} 
\sigma_R^\pm (X; P) \right]_{p^0 = \pm 
\omega_\pm^n (X; \pm {\bf p})} = 0 . 
\label{D2} 
\end{eqnarray} 
%%%%%%%%%%%%%%%%%%%%%%
It is well known \cite{le-b} that, in equilibrium quark-gluon 
plasma, \lq\lq abnormal modes'' called plasmino appears for soft 
$p$ $(= O (g T))$ [$g$ the QCD coupling constant and $T$ the 
temperature]. The energy-shell of such modes, if any, is defined 
through 
%%%%%%%%%%%%%%%%%%%%%%
\begin{eqnarray} 
&& \mbox{Re} \left[ g^\pm_R (X; P) \right]^{- 1}_{p^0 = \mp 
\omega_\pm^a (X; \mp {\bf p})} \nonumber \\ 
&& \mbox{\hspace*{5ex}} 
= \left[ p^0 \mp p \mp \mbox{Re} 
\sigma_R^\pm (X; P) \right]_{p^0 = \mp 
\omega_\pm^a (X; \mp {\bf p})} = 0 . 
\label{D3} 
\end{eqnarray} 
%%%%%%%%%%%%%%%%%%%%%%
Useful formulae that hold on the energy-shell are displayed in 
Appendix C.  

In order to obtain the expression for physical number densities, we 
start with computing charge and momentum densities, 
%%%%%%%%%%%%%%%%%%%%%%%%%%%%%%%%%%%%%%
\begin{eqnarray} 
\mbox{Tr} \left[ j^0 (x) \rho \right] & = & \mbox{Tr} \left[ 
\bar{\psi} (x) \gamma^0 \psi (x) \, \rho \right] \nonumber \\ 
& = & - \frac{i}{2} \mbox{Tr} \left[ \gamma^0 \left( G_{2 1} (x, x) 
+ G_{1 2} (x, x) \right) \rho \right] , \nonumber \\ 
\label{charge} \\ 
\mbox{Tr} \left[ \vec{\cal P} (x) \rho \right] & = & - i \mbox{Tr} 
\left[ \psi^\dagger (x) \nabla \psi (x) \, \rho \right] \nonumber \\ 
& = & - \frac{1}{2} \mbox{Tr} \left[ \gamma^0 \nabla \left( G_{2 1} 
(x, x) + G_{1 2} (x, x) \right) \, \rho \right] . \nonumber \\ 
\label{mom} 
\end{eqnarray} 
%%%%%%%%%%%%%%%%%%%%%%%%%%%%%%%%%%%%%%

We first compute the leading contribution to $\mbox{Tr} \left[ j^0 
(x) \rho \right]$. Substituting the leading parts of $G_{2 1}$ and 
of $G_{1 2}$ (cf. Eq.~(\ref{3.111})) and using Eq.~(\ref{RA0}), we 
obtain 
%%%%%%%%%%%%%%%%%%%%%%%%%%%%%%%%%%%%%%%%%%%%%%
\begin{eqnarray*} 
\mbox{Tr} \left[ j^0 (x) \rho \right] & = & i \int \frac{d^{\, 4} 
P}{(2 \pi)^4} \left[ \left( g_+^R (x; P) + g_-^R (x; P) \right) - 
\mbox{c.c.} \right] \nonumber \\ 
&& \mbox{\hspace*{15ex}} \times [2 f (x; P) - 1] . 
\end{eqnarray*} 
%%%%%%%%%%%%%%%%%%%%%%%
From this, we see that $\mbox{Tr} \left[ j^0 (x) \rho \right]$ 
weakly depends on $x$. The narrow-width approximation, $Im 
\sigma_\pm^R \to \mp i 0^+$ yields 
\widetext 
%%%%%%%%%%%%%%%%%%%%%%%%%%%%%%%%%%
\begin{eqnarray} 
\mbox{Tr} \left[ j^0 (x) \rho \right] & \simeq & 2 \int 
\frac{d^{\, 3} p}{(2 \pi)^3} \left[ Z_+^n (x; \omega_+^n ({\bf p}), 
{\bf p}) n (x; \omega_+^n ({\bf p}), {\bf p}) - Z_-^n (x; \omega_-^n 
({\bf p}), {\bf p}) \bar{n} (x; \omega_-^n ({\bf p}), {\bf p}) 
\right. \nonumber \\ 
&& \left. + Z_-^a (x; \omega_-^a ({\bf p}), {\bf p}) n (x; 
\omega_-^a ({\bf p}), {\bf p}) - Z_+^a (x; \omega_+^a ({\bf p}), 
{\bf p}) \bar{n} (x; \omega_+^a ({\bf p}), {\bf p}) \right] + ... . 
\label{dennka} 
\end{eqnarray}
%%%%%%%%%%%%%%%%%%%%%%%
Here $n$ and $\bar{n}$ are as in Eq.~(\ref{N-desu}), and \lq $...$' 
stands for the contribution from $2 f - 1 \ni - \epsilon (p^0)$ [ 
cf. Eq.~(\ref{f-desu})], which is the vacuum-theory contribution 
corrected by the medium effect. $Z$'s in Eqs.~(\ref{dennka}) are the 
wave-function renormalization factors, Eqs.~(\ref{Z1}) and 
(\ref{D66}) in Appendix C. The first (last) two contributions on the 
RHS come from the \lq\lq normal modes'' (\lq\lq abnormal modes'') of 
quasiparticles. If there are several normal and/or abnormal modes, 
summation should be taken over all modes. The factor \lq $2$' in 
Eqs.~(\ref{dennka}) comes from the spin degrees of freedom. 

In a similar manner, we obtain for the momentum density, 
%%%%%%%%%%%%%%%%%%%%%%%%%%%%%%%%%%
\begin{eqnarray} 
\mbox{Tr} \left[ \vec{\cal P} (x) \rho \right] & \simeq & 2 \int 
\frac{d^{\, 3} p}{(2 \pi)^3} \, {\bf p} \left[ Z_+^n (x; \omega_+^n 
({\bf p}), {\bf p}) n (x; \omega_+^n ({\bf p}), {\bf p}) + Z_-^n 
(x; \omega_-^n ({\bf p}), {\bf p}) \bar{n} (x; \omega_-^n ({\bf p}), 
{\bf p}) \right. \nonumber \\ 
&& \left. + Z_-^a (x; \omega_-^a ({\bf p}), {\bf p}) n (x; 
\omega_-^a ({\bf p}), {\bf p}) + Z_+^a (x; \omega_+^a ({\bf p}), 
{\bf p}) \bar{n} (x; \omega_+^a ({\bf p}), {\bf p}) \right] + ... . 
\label{undo} 
\end{eqnarray}
%%%%%%%%%%%%%%%%%%%%%%%
\narrowtext 

\noindent 
From Eqs.~(\ref{dennka}) and (\ref{undo}), we can read off that 
$n$ is the number density of fermionic quasiparticle and 
$\bar{n}$ the number density of antifermionic quasiparticle. 
Undoing the narrow-width approximation yields further corrections to 
the physical number densities. 

Let us turn to analyze the contributions from the nonleading part of 
$\hat{G}$ in Eq.~(\ref{3.111}). Inspection of Eqs.~(\ref{charge}) 
and (\ref{mom}) with Eqs.~(\ref{3.111})~-~(\ref{GK2}) and 
(\ref{saigo}) shows that all but $G_K^{(0)}$, Eq.~(\ref{GK0}) yield 
well-defined corrections to the physical number densities due to the 
medium effect. $G_K^{(0)}$ contains 
%%%%%%%%%%%%%%%%%%%%%%%%%%%%%%%%%%%%
\[ 
g^\pm_R g^\pm_A = \frac{1}{ \left[ p^0 \mp p \mp \sigma_R^\pm 
\right] \left[ p^0 \mp p \mp \left(\sigma_R^\pm \right)^* \right]} 
. 
\] 
%%%%%%%%%%%%%%%%%%%%%%%%%%%%%%%%%%%%
In the narrow-width approximation $\mbox{Im} \sigma_\pm^R \to 
\mp 0^+$, $g_\pm^R g_\pm^A$ develops pinch singularity in a complex 
$p^0$-plane. Then the contributions of $G_K^{(0)}$ to 
Eqs.~(\ref{charge}) and (\ref{mom}) diverge in this approximation. 
In practice, $\mbox{Im} \sigma_R^\pm$ $(\propto g^2)$ is a small 
quantity, so that the contribution, although not divergent, is 
large. This invalidates the perturbative scheme and a sort of 
\lq\lq renormalization'' is necessary for the number densities 
\cite{nie}. This observation leads us to introduce the condition 
$G_K^{(0)} = 0$ on the energy-shells: 
%%%%%%%%%%%%%%%%%%%%%%%%%%%%%%%%%%%%%%%%%%%%%%%%%%%%%%%%%%
\begin{eqnarray} 
&& \left[ \tilde{P}_\pm \cdot \partial_X f \mp i \left( \sigma^\pm_K 
\right)^{\mbox{\scriptsize{loop}}} \pm \left\{ \mbox{Re} 
\sigma^\pm_R, \; f \right\}_{\mbox{\scriptsize{P.B.}}} 
\right]_{\mbox{\scriptsize{on energy shells}}} \nonumber \\ 
&& \mbox{\hspace*{5ex}} = 0 , 
\label{kiso} 
\end{eqnarray} 
%%%%%%%%%%%%%%%%%%%%%%%%%%%%%
where $\left\{ ... , \; ... \right\}_{\mbox{\scriptsize{P.B.}}}$ is 
as in Eq.~(\ref{poi}). This serves as determining equation for so 
far arbitrary $f$. (See below, for more details.) Now the 
above-mentioned large contributions, which turn out to the diverging 
contributions (due to the pinch singularities) in the narrow-width 
approximation, do not appear. Thus, the contributions from 
$G_K^{(0)}$ to Eqs.~(\ref{charge}) and to (\ref{mom}) also yields 
well-defined corrections to the physical number densities. 
%%%%%%%%%%%%%%%%%%%%%%%%%%%%%%%%%%%%%%%%%%%%%%%
%%%%%% SUB %%%%%%%%%%%%%%%%%%%%%%%%%%%%%%%%%%%%
%%%%%%%%%%%%%%%%%%%%%%%%%%%%%%%%%%%%%
\subsection{Generalized Boltzmann equation} 
We are now in a position to disclose the physical meaning of 
Eq.~(\ref{kiso}). Taking up Eq.~(\ref{kiso}) with upper signs and 
setting $p^0$ on the energy-shell of the normal mode, $p^0 = 
\omega_+^n (X; {\bf p})$, we obtain, using the formulae in Appendix 
C, 
%%%%%%%%%%%%%%%%%%%%%%%%%%%%%%%%%%%%
\begin{eqnarray*} 
&& \left( Z_+^n (X; P) \right)^{- 1} \left[ \frac{\partial}{\partial 
X^0} + {\bf v}_+^n (X; {\bf p}) \cdot \frac{\partial}{\partial {\bf 
X}} \right] n (X; P) \nonumber \\ 
&& \mbox{\hspace*{5ex}} + \frac{\partial \mbox{Re} \sigma^+_R (X; 
P)}{\partial X^\mu} \frac{\partial n (X; P)}{\partial P_\mu} = i 
\left( \sigma^+_K \right)^{\mbox{\scriptsize{loop}}} (X; P) . 
\nonumber 
\end{eqnarray*} 
%%%%%%%%%%%%%%%%%%%%%%%%%%%%%%%%%%%%
Here ${\bf v}_+^n$ $(\equiv \partial \omega_+^n (X; {\bf p}) / 
\partial {\bf p})$ is the group velocity of the mode [cf. 
Eq.~(\ref{bui})]. As will be shown in Appendix D, $i \left( 
\sigma^+_K \right)^{\mbox{\scriptsize{loop}}}$ on the RHS is related 
to the net production rate, $\Gamma^n_{\mbox{\scriptsize{net p}}}$, 
of the mode $p^0 = \omega^n_+ (X; {\bf p})$. Using 
Eqs.~(\ref{sei11}) and (\ref{ato}), we obtain 
%%%%%%%%%%%%%%%%%%%%%%%%%%%%%%%%%%%%
\begin{eqnarray*} 
&& \left[ \frac{\partial}{\partial X^0} + {\bf v}_+^n (X; {\bf p}) 
\cdot \frac{\partial}{\partial {\bf X}} \right] N (X; P) 
\nonumber \\ 
&& \mbox{\hspace*{5ex}} 
+ \frac{\partial 
\omega_+^n (X; {\bf p})}{\partial X^\mu} \frac{\partial N(X; 
P)}{\partial P_\mu} = \Gamma_{\mbox{\scriptsize{net}} \; p}^n (X; 
{\bf p}) . 
\end{eqnarray*} 
%%%%%%%%%%%%%%%%%%%%%%%%%%%%%%%%%%%%
This can further be rewritten in the form, 
%%%%%%%%%%%%%%%%%%%%%%%%%%%%%%%%%%%%
\begin{eqnarray} 
&& \left( \frac{d}{d X^0} + {\bf v}_+^n (X; {\bf p}) \cdot 
\frac{d}{d {\bf X}} \right) n (X; \omega_+^n (X; {\bf p}), \hat{\bf 
p}) 
\nonumber \\ 
&& \mbox{\hspace*{5ex}} - \frac{\partial \omega_+^n (X; {\bf 
p})}{\partial {\bf X}} \frac{d n}{d {\bf P}} = 
\Gamma_{\mbox{\scriptsize{net}} \; p}^n (X; {\bf p}) . 
\label{Bol1} 
\end{eqnarray} 
%%%%%%%%%%%%%%%%%%%%%%%%%%%%%%%%%%%%
Similarly, Eq.~(\ref{kiso}) (with upper signs) with $p_0 = - 
\omega_+^a (X; - {\bf p})$ yields 
%%%%%%%%%%%%%%%%%%%%%%%%%%%%%%%%%%%%
\begin{eqnarray} 
&& \left( \frac{d}{d X^0} + {\bf v}_+^a (X; {\bf p}) \cdot 
\frac{d}{d {\bf X}} \right) \bar{n} (X; \omega_+^a (X; {\bf p}), 
{\bf p}) 
\nonumber \\ 
&& \mbox{\hspace*{5ex}} - \frac{\partial \omega_+^a (X; {\bf 
p})}{\partial {\bf X}} \frac{d \bar{n}}{d {\bf P}} = 
\bar{\Gamma}_{\mbox{\scriptsize{net}} \; p}^a (X; {\bf p}) . 
\label{Bol2} 
\end{eqnarray} 
%%%%%%%%%%%%%%%%%%%%%%%%%%%%%%%%%%%%
Equation~(\ref{kiso}) with lower signs yields, on the energy-shell 
$p_0 = - \omega_-^n (X; - {\bf p})$, 
%%%%%%%%%%%%%%%%%%%%%%%%%%%%%%%%%%%%
\begin{eqnarray} 
&& \left( \frac{d}{d X^0} + {\bf v}_-^n (X; {\bf p}) \cdot 
\frac{d}{d {\bf X}} \right) \bar{n} (X; \omega_-^n (X; {\bf p}), 
{\bf p}) 
\nonumber \\ 
&& \mbox{\hspace*{5ex}} - \frac{\partial \omega_-^n (X; {\bf 
p})}{\partial {\bf X}} \frac{d \bar{n}}{d {\bf P}} = 
\Gamma_{\mbox{\scriptsize{net}} \; p}^n (X; {\bf p}) , 
\label{Bol3} 
\end{eqnarray} 
%%%%%%%%%%%%%%%%%%%%%%%%%%%%%%%%%%%%
and on the energy-shell $p_0 = \omega_-^a (X; {\bf p})$, 
%%%%%%%%%%%%%%%%%%%%%%%%%%%%%%%%%%%%
\begin{eqnarray} 
&& \left( \frac{d}{d X^0} + {\bf v}_-^a (X; {\bf p}) \cdot 
\frac{d}{d {\bf X}} \right) n (X; \omega_-^a (X; {\bf p}), {\bf p}) 
\nonumber \\ 
&& \mbox{\hspace*{5ex}} - \frac{\partial \omega_-^a (X; {\bf 
p})}{\partial {\bf X}} \frac{d n}{d {\bf P}} = 
\Gamma_{\mbox{\scriptsize{net}} \; p}^a (X; {\bf p}) . 
\label{Bol4} 
\end{eqnarray} 
%%%%%%%%%%%%%%%%%%%%%%%%%%%%%%%%%%%%
Equations~(\ref{Bol1})~-~(\ref{Bol4}) are the generalized 
relativistic Boltzmann equation for (anti)fer\-mi\-onic 
quasiparticles. 
%%%%%%%%%%%%%%%%%%%%%%%%%%%%%%%%%%%%%%%%%%%%%%
%%%% SECTION III %%%%%%%%%%%%%%%%%%%%%%%%%%%%%%%
%%%%%%%%%%%%%%%%%%%%%%%%%%%%%%%%%%%%%%%%%%%%%%
%\narrowtext 
\setcounter{equation}{0}
\setcounter{section}{5}
\def\theequation{\mbox{\arabic{section}.\arabic{equation}}} 
\section{Perturbation theory} 
As has been discussed in the preceding section, the propagator in 
the physical-$N$ scheme is free from the pinch singular term (in the 
narrow-width approximation) and then the perturbative calculation of 
some quantity yields \lq\lq healthy'' perturbative corrections. For 
constructing a concrete perturbative scheme, one more step is 
necessary. 

To extend the condition (\ref{kiso}) to off the energy-shell, we 
divide $f$ into two pieces [cf. Eq.~(\ref{f-desu})] 
%%%%%%%%%%%%%%%%%%%%%%%%%%%%%%%%%%%%
\begin{eqnarray} 
f (X; P) & = & \theta (- p^0) + \epsilon (p^0) N (X; P) \nonumber \\ 
& = & \tilde{f} 
(X; P) + f_0 (X; P) \nonumber \\ 
& = & [\theta (- p^0) + \epsilon (p^0) \tilde{N} (X; P)] + \epsilon 
(p^0) N_0 (X; P) . \nonumber \\ 
\label{div} 
\end{eqnarray} 
%%%%%%%%%%%%%%%%%%%%%%%%%%%%%%%%%%%%
$f_0$ (and then also $\tilde{f}$) is defined as follows: Let ${\cal 
R}_i (X; {\bf p})$ $(i = 1, 2, ...)$ be a region in a $p^0$-plane 
that includes $i$th energy-shell. We choose ${\cal R}_i (X; {\bf 
p})$, such that, for $i \neq j$, ${\cal R}_i \cap {\cal R}_j = 
\emptyset$.  On each energy-shell, $N_0 (X; P) = N (X; P)$, and, in 
whole $p^0$-region but ${\cal R}_i (X; {\bf p})$ $(i = 1, 2, ...)$, 
$N_0 (X; P)$ vanishes. $\partial N_0 (X; P) / 
\partial X$ and $\partial N_0 (X; P) / \partial P$ exist and $N_0 
(X; P)$ obeys 
%%%%%%%%%%%%%%%%%%%%%%%%%%%%%%%%%%%%%%%%%%%%%%%%%%%%%%%%%%
\begin{equation} 
\tilde{P}_\pm \cdot \partial_X N_0 \mp i \left( \sigma^\pm_K 
\right)^{\mbox{\scriptsize{{loop}}}} \pm \left\{ \mbox{Re} 
\sigma^\pm_R, \; N_0 \right\}_{\mbox{\scriptsize{P.B.}}} = 0 . 
\label{kiso1} 
\end{equation} 
%%%%%%%%%%%%%%%%%%%%%%%%%%%%%
Then, $G_K^{(0)}$ is given by Eq.~(\ref{GK0}) with $\tilde{f}$ for 
$f$. It is obvious from the above construction that this $G_K^{(0)}$ 
does not possess pinch singularities in narrow-width approximation, 
and thus \lq\lq healthy'' perturbation theory is established. 

It is worth mentioning that there is arbitrariness in the choice of 
the regions ${\cal ,R}_i (X; {\bf p})$ $(i = 1, 2, ...)$. 
Furthermore, the 
choice of the functional forms of $\tilde{f}$ and of $f_0$ is also 
arbitrary, provided that 
%%%%%%%%%%%%%%%%%%%%%%%%%%%%%%%%%%%
\begin{eqnarray*} 
&& \tilde{f} (X^0 = X^0_{in}, {\bf X}; P) + f_0 (X^0= X^0_{in}, {\bf 
X}; P) \\ 
&& \mbox{\hspace*{5ex}} = f (X^0 = X^0_{in}, {\bf X}; P) , 
\end{eqnarray*} 
%%%%%%%%%%%%%%%%%%%%%%%%%%%%%%% 
where $f (X^0 = X^0_{in}, {\bf X}; P)$ is the initial data with 
$X^0_{in}$ the initial time. As has been discussed at the end of 
Sec.~III, these arbitrariness are not the matter. 

To summarize, to the gradient approximation the (resummed) 
propagator $\hat{\cal G}$ of the theory is 
%%%%%%%%%%%%%%%%%%%%%%%%%%%%%
\begin{equation} 
\hat{\cal G} = \hat{G} + \hat{S}_1 + \hat{S}_2 + \hat{S}_3 . 
\label{owao} 
\end{equation} 
%%%%%%%%%%%%%%%%%%%%%%%%%%%%%%
Here $\hat{S}_1$~-~$\hat{S}_3$ are as in Eqs.~(\ref{s1}), 
(\ref{2.66})~-~(\ref{dai1}), and $\hat{G}$ is as in 
Eq.~(\ref{3.111}) provided that $G_K^{(0)}$ is given by 
Eq.~(\ref{GK0}) with $\tilde{f}$ for $f$. $f$ consists of two pieces 
as in Eq.~(\ref{div}). $f_0$ $(= \epsilon (p^0) N_0)$ subjects to 
Eq.~(\ref{kiso1}), which is to be solved under a given initial data. 
It is to be noted that $\hat{S}_1$~-~$\hat{S}_3$ are the nonleading 
parts, so that, if one wants, for $f$ in $\hat{S}_1$~-~$\hat{S}_3$, 
one can substitute $f_B$, the solution to the \lq\lq free Boltzmann 
equation,'' Eq.~(\ref{ba1}). 

As in \cite{nie2}, determination of $f$ or $N$ proceeds order by 
order in perturbation theory, which we do not reproduce here. 

The vertex factor and the initial correlations are the same as in 
standard CTP formalism, except that an additional two-point vertex 
$i L_c (x, y) \hat{A}_-$, Eq.~(\ref{sigmac}), exists. It is to be 
noted that the two-point vertex $i L_c (x, y) \hat{A}_-$ has been 
built into $\hat{G}$ (cf. Eqs.~(\ref{GKkota}) and (\ref{GK0}) with 
Eqs.~(\ref{GK}) and (\ref{56})) and is absent in the perturbative 
framework using $\hat{\cal G}$ in Eq.~(\ref{owao}). 
%%%%%%%%%%%%%%%%%%%%%%%%%%%%%%%%%%%%%%%%%%%%%%
%%%% SECTION III %%%%%%%%%%%%%%%%%%%%%%%%%%%%%%%
%%%%%%%%%%%%%%%%%%%%%%%%%%%%%%%%%%%%%%%%%%%%%%
\setcounter{equation}{0}
\setcounter{section}{6}
\def\theequation{\mbox{\arabic{section}.\arabic{equation}}} 
\section{Summary and discussion} 
In this paper we have dealt with out-of-equilibrium perturbation 
theory for massless Dirac fermions. The fermion propagator is 
constructed from first principles. Essentially only approximation 
we have employed is the so-called gradient approximation, so that 
the perturbative framework applies to the quasiuniform systems near 
equilibrium or the nonequilibrium quasistationary systems. The 
framework allows us to compute any reaction rates \cite{noo}. 

There comes out naturally the generalized Boltzmann equation (GBE) 
that describes the spacetime evolution of the number densities of 
quasiparticles, through which the evolution of the system is 
described. 

Comparison of the present formalism with some earlier works has been 
made in \cite{nie}. We like to add here two related works 
\cite{hen,B-I}, in which fermions are dealt with. In these papers, 
the GBE is derived in a traditional manner by starting with the SD 
equation (cf. Eq.~(\ref{SD})). As is mentioned in \cite{nie}, the SD 
equation is nothing more than an equation that serves as resumming 
the self-energy part to makeup the resummed propagator, 
Eq.~(\ref{3.111}). Then, in order to derive the GBE, an additional 
input or condition is necessary. Our condition is Eq.~(\ref{kiso}). 
The additional input in \cite{hen,B-I} is, in our notation, an 
introduction of $f'$ through 
%%%%%%%%%%%%%%%%%%%%%%%%%
\begin{eqnarray}
G_{1 2 (2 1)} (X; P) & \simeq & 2 \pi i f' (X; P) \epsilon (p_0) 
{P\kern-0.1em\raise0.3ex\llap{/}\kern0.15em\relax} \sigma (P) 
\nonumber \\ 
& = & 2 \pi i \left[ \theta (p_0) N' (X; P) \right. \nonumber \\ 
&& \left. \mbox{\hspace*{4ex}} - \theta (- p_0) \left( 1 - \bar{N}' 
(X; P) \right) \right] 
{P\kern-0.1em\raise0.3ex\llap{/}\kern0.15em\relax} \sigma (P) 
\nonumber \\ 
\label{ans} 
\end{eqnarray}
%%%%%%%%%%%%%%%%%%%%%%
with $\sigma (P) \simeq \delta (P^2)$. In \cite{hen}, this is done 
on the basis of the quasiparticle picture, and, in \cite{B-I}, on 
the order-of-magnitude estimation of $\hat{\Sigma}$ in diagrammatic 
analysis. Substituting Eq.~(\ref{ans}) into the Kadanoff-Baym 
equation, which is a part of the SD equation, the GBE for $N' (X; 
P)$ and $\bar{N}' (X; P)$ results. In contrast to the formalism 
in the present paper, no counter Lagrangian is explicitly 
introduced there and then the consistency check for the formalism 
seems to be necessary. 

In \cite{nie,nie1}, comparison has been made of the present 
formalism with nonequilibrium thermo field dynamics (NETFD) 
\cite{ume}. (See also \cite{hen}). We like to recapitulate here 
that, in NETFD, a counter action is introduced on the basis of the 
renormalizability argument. Imposition of the renormalization 
condition leads to the GBE. 

The \lq\lq derivation'' in this paper of the GBE for a quark-gluon 
plasma (nonequilibrium QCD) is quite different from the traditional 
derivation (cf. \cite{hen,B-I}). What we have shown here is that the 
requirement of the absence of large contributions from the 
perturbative framework leads to the GBE. This means that the 
quasiparticles thus defined are the well-defined modes in the 
medium. Conversely, if we start with defining the quasiparticles 
such that their number density functions subject to the GBE, then, 
on the basis of them, well-defined perturbation theory may be 
constructed. In \cite{ume}, the GBE is derived by imposing the 
renormalization condition'' for the propagator. Since the pinch 
singularity, which arises in the narrow-width limit in our 
formulation, is a singularity in momentum space, it is not 
immediately obvious how to translate this condition into (space)time 
representation, as adopted in \cite{ume}. Nevertheless, closer 
inspection of the structures of our formalism and of the NETFD tells 
us that our condition is in accord with on-shell renormalization 
condition in NETFD. 
%%%%%%%%%%%%%%%%%%%%%%%%%%%%
\section*{Acknowledgments}
This work was supported in part by a Grant-in-Aide for Scientific 
Research ((C)(2) (No.~12640287)) of the Ministry of Education, 
Science, Sports and Culture of Japan. 
%%%%%%%%%%%%%%%%%%%%%%%%%%%%%%%
\begin{appendix} %%%%%%%%%%%%%%%%%%%%%%%%%%%%%%%
%%%% APP %%%%%%%%%%%%%%%%%%%%%%%%%%%%%%%%%%%%%%%%%%%%%%%%%%%
%%%%%%%%%%%%%%%%%%%%%%%%%%%%%%%%%%%%%%%%%%%%%%%%%%%%%%%%%%%%
\setcounter{equation}{0}
\setcounter{section}{0}
\section{Expression for the nonleading part of $\hat{S} (X; P)$} 
\def\theequation{\mbox{\Alph{section}.\arabic{equation}}}
The nonleading parts, $\hat{S}_2$ and $\hat{S}_3$, of $\hat{S}$ 
read 
\widetext 
%%%%%%%%%%%%%%%%%%%%%%%%%
\begin{eqnarray} 
\hat{S}_2 (X; P) & = & - \left[ \Delta_R (P) - \Delta_A (P) 
\right] \epsilon (p^0) \gamma_5 
{P\kern-0.1em\raise0.3ex\llap{/}\kern0.15em\relax} N^{(-)} (P) 
\hat{A}_+ , 
\label{2.66} \\ 
\hat{S}_3 (X; P) & = & - \left[ \Delta_R (P) - \Delta_A (P) 
\right] \epsilon (p^0) \left[ \left\{ \sigma^{2 j} p^j + 
\epsilon^{1 i j} \sigma^{0 i} p^j - \frac{\sigma^{3 j} p^2 p^j + 
\epsilon^{3 i j} \sigma^{0 i} p^1 p^j}{p^0 + p^3} \right\} \mbox{Re} 
{\cal 
N} (P) \right. \nonumber \\ 
&& + \left. \left\{ \sigma^{1 j} p^j - \epsilon^{2 i j} \sigma^{0 
i} p^j - \frac{\sigma^{3 j} p^1 p^j - \epsilon^{3 i j} \sigma^{0 i} 
p^2 p^j}{p^0 + p^3} \right\} \mbox{Im} 
{\cal N} (P) \right] \hat{A}_+ , 
\label{dai1} 
\end{eqnarray} 
%%%%%%%%%%%%%%%%%%%%%%%%%%%%%%%%%%%%%%%%%%%%%%
where $\epsilon^{123} = 1$, $\sigma^{\mu \nu} = i (\gamma^\nu 
\gamma^\mu - \gamma^\mu \gamma^\nu) / 2$, and 
%%%%%%%%%%%%%%%%%%%%%%%%%%%%%%%%%%%%%%%%%%%%%%%%%%%%%%%%%%%%%
\begin{eqnarray} 
N^{(-)} (X; P) & \equiv & \frac{1}{2} \left[ 
\theta (p^0) \left\{ N_{+ +} (X; |p^0|, \tilde{\bf p}) - N_{- -} (X; 
|p^0|, \tilde{\bf p}) \right\} \right. \nonumber \\ 
&& \left. - \theta (- p^0) \left\{ \bar{N}_{++} (X; |p^0|, - 
\tilde{\bf p}) - \bar{N}_{- -} (X; |p^0|, - \tilde{\bf p}) \right\} 
\right] \nonumber \\ 
& \simeq & N^{(-)} (P) , 
\nonumber \\ 
{\cal N} (P) & \equiv & \theta (p^0) N_{- +} (|p_0|, \tilde{\bf p}) 
+ \theta (- p^0) \bar{N}_{- +} (|p_0|, - \tilde{\bf p}) . \nonumber 
% \label{yato} 
\end{eqnarray} 
%%%%%%%%%%%%%%%%%%%%%%%%%%%%%%%%%%%%%%%%%%%%%%
\narrowtext 

\noindent 
As mentioned at the beginning of Sec.~IIB, we have assumed that 
$\hat{S}_2$ and $\hat{S}_3$ are, at most, of the same order of 
magnitude as the gradient term $\hat{S}_1$, and then $X$-dependence 
of $N^{(-)}$ and ${\cal N}$ have been ignored. It can readily be 
seen that, to the gradient approximation, 
$i \hat{\tau}_3 
{\partial\kern-0.1em\raise0.3ex\llap{/}\kern0.15em\relax}_x 
\hat{S}_2 (x, y) 
= i \hat{\tau}_3 
{\partial\kern-0.1em\raise0.3ex\llap{/}\kern0.15em\relax}_x 
\hat{S}_3 (x, y) = 0$ holds. One can see from Eq.~(\ref{dai1}) that 
$\hat{S}_3$ does not contribute to the unpolarized quantities. 

For a charge-conjugation-invariant system, $N^{(-)} (P) 
= N^{(-)} (- P)$ and ${\cal N} (P) = {\cal N} (- P)$ hold. 
%%%%%%%%%%%%%%%%%%%%%%%%%%%%%%%%%%%%%%
%%%% APP %%%%%%%%%%%%%%%%%%%%%%%%%%%%%%%%%%%%%%%%%%%%%%%%%%%
%%%%%%%%%%%%%%%%%%%%%%%%%%%%%%%%%%%%%%%%%%%%%%%%%%%%%%%%%%%%
\setcounter{equation}{0}
\setcounter{section}{1}
\def\theequation{\mbox{\Alph{section}.\arabic{equation}}}
\section{Solving the SD equation, Eq.~(\ref{SD})} 
Multiplying $\hat{B}_L^{- 1} \cdot = \hat{\tau}_3 \hat{B}_R 
\hat{\tau}_3 \cdot$ ($\cdot \hat{B}_R^{- 1} = \cdot \hat{\tau}_3 
\hat{B}_L \hat{\tau}_3)$ [cf. Eq.~(\ref{4.111})] from the left 
(right) of each term in Eq.~(\ref{SD}), we obtain 
%%%%%%%%%%%%%%%%%%%%%%%%%%%%%%%%%%%%%%%%%%%%%%%%%%%%%%%%%%%%%%%%
\begin{eqnarray} 
\underline{\hat{G}} & = & \underline{\hat{\cal S}} + 
\underline{\hat{\cal S}} \cdot \underline{\hat{\Sigma}} \cdot 
\underline{\hat{G}} = \underline{\hat{\cal S}} + \underline{\hat{G}} 
\cdot \underline{\hat{\Sigma}} \cdot \underline{\hat{\cal S}} , 
\label{hos} \\ 
\underline{\hat{G}} & \equiv & \hat{B}_L^{- 1} \cdot \hat{G} 
\cdot \hat{B}_R^{- 1} , 
\label{hen} 
\\ 
\underline{\hat{\cal S}} & \equiv & \hat{B}_L^{- 1} \cdot 
\hat{\cal S} \cdot \hat{B}_R^{- 1} = \left( 
\begin{array}{cc} 
i {\partial\kern-0.1em\raise0.3ex\llap{/}\kern0.15em\relax} 
\Delta_R & \;\;\, S_K \\ 
0 & \;\;\, - i 
{\partial\kern-0.1em\raise0.3ex\llap{/}\kern0.15em\relax} 
\Delta_A 
\end{array} 
\right) , 
\label{sb} 
\\ 
\underline{\hat{\Sigma}} & \equiv & \hat{B}_R \cdot \hat{\Sigma} 
\cdot \hat{B}_L = \left( 
\begin{array}{cc} 
\Sigma_R & \;\;\, \Sigma_K \\ 
0 & \;\;\, - \Sigma_A 
\end{array} 
\right) . 
\label{sigb} 
\end{eqnarray} 
%%%%%%%%%%%%%%%%%%%%%%%%%%%%
Here $\Sigma_{R (A)} = \Sigma_{1 1} + \Sigma_{1 2 (2 1)}$, and $S_K$ 
and $\Sigma_K$ are the inverse Wigner transforms of 
%%%%%%%%%%%%%%%%%%%%%%%%%%%%%%%%%%%%%%%%%%%%%%%%%%%%%%%%%%
\begin{eqnarray} 
S_K (X; P) & = & i \left[ 
{\partial\mbox{\hspace*{0.3mm}}\kern-0.1em\raise0.3ex\llap{/}
\kern0.15em\relax}_X f (X; P) \right. \nonumber \\ 
&& \left. \mbox{\hspace*{2ex}} - \frac{\gamma^0 + \gamma^3}{p^0 + 
p^3} P \cdot \partial_X f (X; P) \right] \frac{\bf P}{P^2} , 
\nonumber \\ 
\label{SK} 
\end{eqnarray} 
%%%%%%%%%%%%%%%%%%%%%%%%%%%%%%%%%%%%%%%%
and of 
%%%%%%%%%%%%%%%%%%%%%%%%%%%%%%%%%%%%%%%%%%%%%%%%%%%%%%%%%%
\begin{eqnarray} 
\Sigma_K (X; P) & = & [ 1 - f (X; P)] \Sigma_{1 2} (X; P) \nonumber 
\\ 
&& + f (X; P) \Sigma_{2 1} (X; P) \nonumber \\ 
&& + \frac{i}{2} \left\{ \Sigma_R + \Sigma_A, \; f 
\right\}_{\mbox{\scriptsize{P.B.}}} , 
\label{SigK} 
\end{eqnarray} 
%%%%%%%%%%%%%%%%%%%%%%%%%%%%%%%%%%%%%%%%
respectively. Here $\left\{ ... , \; ... 
\right\}_{\mbox{\scriptsize{P.B.}}}$ is as in Eq.~(\ref{poi}). In 
obtaining Eq.~(\ref{SK}) [Eq.~(\ref{SigK})], use has been made of 
Eq.~(\ref{meme}) [Eq.~(\ref{4.66})]. Although the last term in 
Eq.~(\ref{SigK}) may be dropped to the approximation under 
consideration, we have kept it. 

From the above definitions of $\Sigma_{R (A)}$, $\Sigma_K$, 
and of $\hat{\Sigma}^{(c)}$ (Eq.~(\ref{sigmac})), we have 
%%%%%%%%%%%%%%%%%%%%%%%%%%%%%%%%%%%%%%%%%%%%%
\begin{eqnarray} 
\Sigma_R^{(c)} & = & \Sigma_A^{(c)} = 0 , \nonumber \\ 
\Sigma_K^{(c)} (X; P) & = & i \frac{\gamma^0 + \gamma^3}{p^0 + p^3} 
P \cdot \partial_X f (X; P) . 
\label{56} 
\end{eqnarray} 
%%%%%%%%%%%%%%%%%%%%%%%%%%%%%%%%%%%%%%%%%%%%%

As seen from Eqs.~(\ref{sb}) and (\ref{sigb}), $\hat{\cal S}$ and 
$\hat{\Sigma}$ are triangular matrices, so that Eq.~(\ref{hos}) may 
easily be solved to yield  
%%%%%%%%%%%%%%%%%%%%%%%%%%%%%%%%%%%%%%%%%%%%%
\begin{eqnarray} 
\underline{\hat{G}} & = & \left( 
\begin{array}{cc} 
G_R & \;\;\, G_K' \\ 
0 & \;\;\, - G_A 
\end{array} 
\right) , 
\nonumber 
% label{GRA} 
\\ 
G_{R (A)} & = & \left[ i 
{\partial\kern-0.1em\raise0.3ex\llap{/}\kern0.15em\relax} 
\Delta_{R (A)} - \Sigma_{R (A)} \right]^{- 1} , 
\label{3.101} 
\\ 
G_K' & = & G_R \cdot \left[ \left( i 
{\partial\kern-0.1em\raise0.3ex\llap{/}\kern0.15em\relax} 
\Delta_R \right)^{- 1} \cdot S_K \cdot \left( i 
{\partial\kern-0.1em\raise0.3ex\llap{/}\kern0.15em\relax} \Delta_A 
\right)^{- 1} - \Sigma_K \right] \cdot G_A . \nonumber \\ 
\label{GK} 
\end{eqnarray} 
%%%%%%%%%%%%%%%%%%%%%%%%%%%%%%%%%%%%%%%%%%%%%
Substituting this back into $\hat{G} = \hat{B}_L \cdot 
\underline{\hat{G}} \cdot \hat{B}_R$, Eq.~(\ref{hen}), we obtain, 
after Wigner transformation, Eq.~(\ref{3.111}) with 
%%%%%%%%%%%%%%%%%%%%%%%%%%%%%%%%%%%%%%%%%%%%%
\[ 
G_K (X; P) = G_K' (X; P) + \frac{i}{2} \left\{ G_R + G_A , \; f 
\right\}_{\mbox{\scriptsize{P.B.}}} . 
\] 
%%%%%%%%%%%%%%%%%%%%%%%%%%%%%%%%%%%%%%%%%%%%%
From Eq.~(\ref{3.101}), we obtain, after some manipulation, 
Eq.~(\ref{GRkota}) with Eq.~(\ref{RA0}) and with 
%%%%%%%%%%%%%%%%%%%%%%
\begin{eqnarray} 
G_{R (A)}^{(1)} (X; P) & = & \frac{1}{4 p} \left[ g_{R (A)}^+ (X; P) 
- g_{R (A)}^- (X; P) \right] \nonumber \\ 
&& \times \sum_{\tau = \pm} \left[ \gamma_5 
\vec{\gamma} \cdot \left( \tilde{\bf p} \times \nabla_{X} \sigma_{R 
(A)}^\tau (X; P) \right) \right. \nonumber \\ 
&& \mbox{\hspace*{8ex}} \left. \times g_{R (A)}^\tau (X; P) \right] 
, 
\label{saigo} 
\end{eqnarray} 
%%%%%%%%%%%%%%%%%%%%%%%%%%%%%%%%%%%%
where $g^\pm_{R (A)}$ is as in Eq.~(\ref{4.21}) and $\sigma_{R 
(A)}^\tau = \sigma_{1 1}^\tau + \sigma_{1 2 (2 1)}^\tau$. 
%%%%%%%%%%%%%%%%%%%%%%%%%%%%%%%%%%%%%%%%%%%%%%%%%%%%%%%%%
%%%% APP %%%%%%%%%%%%%%%%%%%%%%%%%%%%%%%%%%%%%%%%%%%%%%%%%%%
%%%%%%%%%%%%%%%%%%%%%%%%%%%%%%%%%%%%%%%%%%%%%%%%%%%%%%%%%%%%
\setcounter{equation}{0}
\setcounter{section}{2}
\section{On the energy shell} 
\def\theequation{\mbox{\Alph{section}.\arabic{equation}}}
Here we display some formulae, which hold on the energy-shells of 
quasiparticles [cf. Eqs.~(\ref{D2}) and (\ref{D3}) with 
Eqs.~(\ref{RA0}) and (\ref{4.21})]. In most formulae in this 
Appendix, we drop the argument $X$. 

\noindent {\em Normal modes}: 

We define the wave-function renormalization factors through taking 
derivative of Eq.~(\ref{D2}) with respect to $p^0$: 
%%%%%%%%%%%%%%%%%%%%%%
\begin{equation} 
\left( Z_\pm^n (\omega_\pm^n ({\bf p}), {\bf p}) \right)^{- 1}= 1 
\mp \frac{\partial \, \mbox{Re} \, 
\sigma_R^{\pm} (p^0, \pm {\bf p})}{\partial 
p^0} \, \rule[-3mm]{.14mm}{8.5mm} 
\raisebox{-2.85mm}{\scriptsize{$\; p^0 = \pm \omega_\pm^n ({\bf 
p})$}} . 
\label{Z1} 
\end{equation} 
%%%%%%%%%%%%%%%%%%%%%%
The group velocities of the modes are obtained from the definition 
(\ref{D2}), 
%%%%%%%%%%%%%%%%%%%%%%
\begin{eqnarray} 
{\bf v}_\pm^n ({\bf p}) & \equiv & \frac{d \omega_\pm^n ({\bf 
p})}{d {\bf p}} \nonumber \\ 
& = & Z_\pm^n (\omega_\pm^n ({\bf p}) , {\bf p}) 
\nonumber \\ 
&& \times \left[ 
\hat{\bf p} + \frac{\partial \, \mbox{Re} \, 
\sigma_R^{\pm} (p^0, \pm {\bf 
p})}{\partial {\bf p}} \, 
\rule[-3mm]{.14mm}{8.5mm} \raisebox{-2.85mm}{\scriptsize{$\; p^0 = 
\pm \omega_\pm^n ({\bf p})$}} \right] . 
\label{bui} 
\end{eqnarray} 
%%%%%%%%%%%%%%%%%%%%%%
By differentiating Eq.~(\ref{D2}) with respect to $X$, we obtain 
%%%%%%%%%%%%%%%%%%%%%%
\begin{equation} 
\frac{\partial \omega_+^n (X; {\bf p})}{\partial X} = Z_+^n 
(\omega_+^n ({\bf p}), {\bf p}) \frac{\partial \, \mbox{Re} \, 
\sigma^+_R (X; \omega_+^n (X; {\bf p}), {\bf p})}{\partial X} . 
\label{ato} 
\end{equation} 
%%%%%%%%%%%%%%%%%%%%%%

\noindent {\em Abnormal modes}: 

%%%%%%%%%%%%%%%%%%%%%%
\begin{eqnarray} 
&& \left( Z_\pm^a (\omega_\pm^a ({\bf p}), {\bf p}) \right)^{- 1} 
= 1 \mp \frac{\partial \, \mbox{Re} \, \sigma_R^{\pm} (p^0, \mp {\bf 
p})}{\partial p^0} \, \rule[-3mm]{.14mm}{8.5mm} 
\raisebox{-2.85mm}{\scriptsize{$\; p^0 = \mp \omega_\pm^a ({\bf 
p})$}} , \nonumber \\ 
\label{D66} \\ 
&& {\bf v}_\pm^a ({\bf p}) \equiv \frac{d \omega_\pm^a ({\bf 
p})}{d {\bf p}} 
\nonumber \\  
&& \mbox{\hspace*{6.3ex}} = - Z_\pm^a (\omega_\pm^a ({\bf p}) , 
{\bf p}) \nonumber \\ 
&& \mbox{\hspace*{12ex}} \times \left[ 
\hat{\bf p} + \frac{\partial \, \mbox{Re} \, 
\sigma_R^{\pm} (p^0, \mp {\bf 
p})}{\partial {\bf p}} \, 
\rule[-3mm]{.14mm}{8.5mm} \raisebox{-2.85mm}{\scriptsize{$\; p^0 = 
\mp \omega_\pm^a ({\bf p})$}} \right] . \nonumber 
\end{eqnarray} 
%%%%%%%%%%%%%%%%%%%%%%
%%%%%%%%%%%%%%%%%%%%%%%%%%%%%%%%%%%%%%%%%%%%%%%%%%%%%%%%%%%%
%%%% APP %%%%%%%%%%%%%%%%%%%%%%%%%%%%%%%%%%%%%%%%%%%%%%%%%%%
%%%%%%%%%%%%%%%%%%%%%%%%%%%%%%%%%%%%%%%%%%%%%%%%%%%
\setcounter{equation}{0}
\setcounter{section}{3}
\section{Net production rates} 
\noindent $p^0 = \omega_\pm^{n / a} ({\bf p})$ {\em fermionic 
modes}: 

From Eqs.~(\ref{RA0}), (\ref{D2}), and (\ref{D3}), we see that the 
projection operators onto $p^0 = \omega_+^n$ $(p^0 = \omega_-^a)$ 
mode is $\tilde{P\kern-0.1em\raise0.3ex\llap{/}\kern0.15em\relax}_+$ 
$(\tilde{P\kern-0.1em\raise0.3ex\llap{/}\kern0.15em\relax}_-)$. 
Then, the production and decay rates are written, in respective 
order, as \cite{chou,nie} 
%%%%%%%%%%%%%%%%%%%%%%%%%%%%%%%%%%%%
\begin{eqnarray*} 
\Gamma_p^{n / a} ({\bf p}) & = & \frac{i}{4} Z_\pm^{n / a} 
(\omega_\pm^{n / a} ({\bf p}), {\bf p}) \nonumber \\ 
&& \times \mbox{Tr} \left[ \left( \Sigma_{1 2} (\omega_\pm^{n / a} 
({\bf p}), {\bf p}) \right)^{\mbox{\scriptsize{loop}}} 
\tilde{P\kern-0.1em\raise0.3ex\llap{/}\kern0.15em\relax}_\pm \right] 
\nonumber \\ 
& = & \pm i Z_\pm^{n / a} \left( \sigma^\pm_{1 2} (\omega_\pm^{n / 
a} ({\bf p}), {\bf p}) \right)^{\mbox{\scriptsize{loop}}} , 
\nonumber \\ 
\Gamma_d^{n / a} (X; {\bf p}) & = & - \frac{i}{4} Z_\pm^{n / a} 
(\omega_\pm^{n / a} ({\bf p}), {\bf p}) \nonumber \\ 
&& \times \mbox{Tr} \left[ 
\left( \Sigma_{2 1} (\omega_\pm^{n / a} ({\bf p}), {\bf p}) 
\right)^{\mbox{\scriptsize{loop}}} 
\tilde{P\kern-0.1em\raise0.3ex\llap{/}\kern0.15em\relax}_\pm 
\right] \nonumber \\ 
& = & \mp i Z_\pm^{n / a} \left( \sigma^\pm_{2 1} (\omega_\pm^{n 
/ a} ({\bf p}), {\bf p}) \right)^{\mbox{\scriptsize{loop}}} , 
% \label{sei1} 
\end{eqnarray*} 
%%%%%%%%%%%%%%%%%%%%%%%%%%%%%%%%%%%%
where $Z$'s are the wave-function renormalization factor, 
Eqs.~(\ref{Z1}) and (\ref{D66}). 
Thus, the net production rate is 
%%%%%%%%%%%%%%%%%%%%%%%%%%%%%%%%%%%%
\begin{eqnarray} 
\Gamma_{\mbox{\scriptsize{net}} \; p}^{n / a} ({\bf p}) & = & [1 - 
n (\omega_\pm^{n / a} 
({\bf p}), \tilde{\bf p})] \Gamma_p^{n / a} (X; {\bf p}) \nonumber 
\\ 
&& - n (\omega_\pm^{n / a} ({\bf p}), \tilde{\bf p}) \Gamma_d^{n / 
a} ({\bf p}) \nonumber \\ 
& = & \pm i Z_\pm^{n / a} (\omega_\pm^{n / a} ({\bf p}), {\bf p}) 
\nonumber \\ 
&& \times \left( \sigma^\pm_K (\omega_\pm^{n / a} (X; {\bf p}), {\bf 
p}) \right)^{\mbox{\scriptsize{loop}}} , 
\label{sei11} 
\end{eqnarray} 
%%%%%%%%%%%%%%%%%%%%%%%%%%%%%%%%%%%%
where $\left( \sigma^\pm_K \right)^{\mbox{\scriptsize{loop}}}$ is 
as in Eq.~(\ref{3.188}). 

\noindent $p^0 = - \omega_\mp^{n / a} (- {\bf p})$ {\em 
antifermionic modes}: 
%%%%%%%%%%%%%%%%%%%%%%%%%%%%%%%%%%%%
\begin{eqnarray*} 
\overline{\Gamma}_p^{n / a} (- {\bf p}) & = & - \frac{i}{4} 
Z_\mp^{n / a} (\omega_\mp^{n / a} (- {\bf p}), - {\bf p}) 
\nonumber \\ 
&& \times \mbox{Tr} \left[ \left( \Sigma_{2 1} (- \omega_\mp^{n / 
a} (- {\bf 
p}), {\bf p}) \right)^{\mbox{\scriptsize{loop}}} 
\tilde{P\kern-0.1em\raise0.3ex\llap{/}\kern0.15em\relax}_\mp 
\right] \nonumber \\ 
& = & \pm i Z_\mp^{n / a} \left( \sigma^\mp_{2 1} (- \omega_\mp^{n 
/ a} (- {\bf p}), {\bf p}) \right)^{\mbox{\scriptsize{loop}}} , 
\nonumber \\ 
\overline{\Gamma}_d^{n / a} (- {\bf p}) & = & \frac{i}{4} Z_\mp^{n / 
a} (\omega_\mp^{n / a} (- {\bf p}), - {\bf p}) \nonumber \\ 
&& \times \mbox{Tr} \left[ \left( \Sigma_{1 2} (- \omega_\mp^{n / a} 
(- {\bf p}) , {\bf p}) \right)^{\mbox{\scriptsize{loop}}} 
\tilde{P\kern-0.1em\raise0.3ex\llap{/}\kern0.15em\relax}_\mp \right] 
\nonumber \\ 
& = & \mp i Z_\mp^{n / a} \left( \sigma^\mp_{1 2} (- \omega_\mp^{n 
/ a} (- {\bf p}), {\bf p}) \right)^{\mbox{\scriptsize{loop}}} , 
\nonumber \\ 
\overline{\Gamma}_{\mbox{\scriptsize{net}} \; p}^{n / a} (- 
{\bf p}) & = & [1 - \bar{n} ( \omega_\mp^{n / a} (- {\bf p}),  - 
{\bf p})] \overline{\Gamma}_p^{n / a} (- {\bf p}) \nonumber \\ 
&& - \bar{n} ( \omega_\mp^{n / 
a} (- {\bf p}), - {\bf p}) \overline{\Gamma}_d^{n / a} (- {\bf p}) 
\nonumber \\ 
& = & \pm i Z_\mp^{n / a} (\omega_\mp^{n / a} (- {\bf p}), - {\bf 
p}) \nonumber \\ 
&& \mbox{\hspace*{3ex}} \times \left( \sigma^\mp_K (- \omega_\mp^{n 
/ a} (- {\bf p}), {\bf p}) 
\right)^{\mbox{\scriptsize{loop}}} . 
\end{eqnarray*} 
%%%%%%%%%%%%%%%%%%%%%%%%%%%%%%%%%%%%
\end{appendix} 

\end{document}